\documentclass{JHEP3}

\usepackage{amsmath}
\usepackage{epsfig,multicol,bbm}
\usepackage{graphicx}
\usepackage{bm}
\usepackage{multirow}
\usepackage{xspace}

\newcommand{\nn}{\nonumber} 
\newcommand{\bn}{{\bar n}}

\newcommand{\be}{\begin{equation}}
\newcommand{\ee}{\end{equation}}

\newcommand{\ket}[1]{\left\lvert #1\right\rangle}

\newcommand{\as}{\alpha_s}

\newcommand{\kt}{\textrm{k}_\textrm{T}}
\newcommand{\akt}{\textrm{ak}_\textrm{T}}

\newcommand{\ca}[1]{\mathcal{#1}}
\newcommand{\ord}[1]{\mathcal{O}(#1)}

\newcommand{\wt}{\widetilde}
\newcommand{\cM}{\wt{\ca{M}}}
\newcommand{\cd}{\wt{w}}
\newcommand{\de}{w}
\newcommand{\R}{\mathcal{R}}
\newcommand{\cset}[1]{(\{#1\})}

\newcommand{\alg}{\text{alg}}

\newcommand{\eq}[1]{Eq.~\eqref{eq:#1}}

\renewcommand{\sec}[1]{Sec.~\ref{sec:#1}}
\newcommand{\ssec}[1]{Sec.~\ref{ssec:#1}}
\newcommand{\fig}[1]{Fig.~\ref{fig:#1}}

\newcommand{\tab}[1]{Table~\ref{tab:#1}}

\allowdisplaybreaks[3]

\title{Disentangling Clustering Effects in Jet Algorithms}

 \author{
Randall Kelley \\
Department of Physics, Harvard University, Cambridge, Massachusetts, 02138, USA \\ 
E-mail: \email{rkelley@physics.harvard.edu}
}

\author{
Jonathan R. Walsh and Saba Zuberi \\
Theoretical Physics Group, Ernest Orlando Lawrence Berkeley National Laboratory, \\
and Center for Theoretical Physics, University of California, Berkeley, CA 94720, USA \\
E-mail: \email{jwalsh@lbl.gov}, \email{szuberi@lbl.gov}
 }

\abstract{Clustering algorithms build jets though the iterative application of single particle and pairwise metrics. This leads to phase space constraints that are extremely complicated beyond the lowest orders in perturbation theory, and in practice they must be implemented numerically. This complication presents a significant barrier to gaining an analytic understanding of the perturbative structure of jet cross sections. We present a novel framework to express the jet algorithm's phase space constraints as a function of clustered groups of particles, which are the possible outcomes of the algorithm. This approach highlights the analytic properties of jet observables, rather than the explicit constraints on individual final state momenta, which can be unwieldy at higher orders.  We derive the form of the $n$-particle phase space constraints for a jet algorithm with any measurement. We provide an expression for the measurement that makes clustering effects manifest and relates them to constraints from clustering at lower orders.  The utility of this framework is demonstrated by using it to understand clustering effects for a large class of jet shape observables in the soft/collinear limit. We apply this framework to isolate divergences and analyze the logarithmic structure of the Abelian terms in the soft function, providing the all-orders form of these terms and showing that corrections from clustering start at next-to-leading logarithmic order in the exponent of the cross section.
}


\begin{document}

%Main body of the paper
%%%%%%%%%%%%%%%%%%%%%%%%%%%%%%%%%%%%%%%%%%%%%%%%%%%
%%%%%%%%%%%%%%%%%%%%%%%%%%%%%%%%%%%%%%%%%%%%%%%%%%%

%%%%%%%%%%%%%%%%%%%%%%%%%%%%%%%%%%%%%%%%%%%%%%%%%%%
\section{Introduction}
\label{sec:intro}
%%%%%%%%%%%%%%%%%%%%%%%%%%%%%%%%%%%%%%%%%%%%%%%%%%%

Jet algorithms play an essential role in high energy collision experiments, organizing the hadronic structure of an event to allow for a meaningful interpretation of the short distance interaction.  The LHC has increased interest in understanding and utilizing the structure of jets \cite{Abdesselam:2010pt,Altheimer:2012mn}, and these tools often make use of clustering algorithms such as anti-$\kt$, Cambridge/Aachen (C/A), and $\kt$ \cite{Catani:1991hj,Catani:1993hr,Ellis:1993tq,Dokshitzer:1997in,Cacciari:2008gp}.  Understanding the structure of jets requires understanding their perturbative description. 

Clustering algorithms build jets through an iterative recombination procedure that sequentially identifies the pair of particles that are closest according to a distance measure and then \emph{clusters} them, i.e.~combines their momenta, doing so until a stopping criterion is reached.  Each algorithm differs mainly in the choices of the distance measure and the stopping criterion used in this procedure.  Implementing phase space constraints from clustering for final states with more than two or three particles is very complicated, and can present an impasse to analytic perturbative calculations, which in many cases must be performed numerically. This makes it difficult to understand the effects of clustering on the perturbative series in jet cross sections.

%------------------------------------------------------------------------------------------------------------------------------------------------
\subsection{The Complication of Measuring Properties of Jets}
%------------------------------------------------------------------------------------------------------------------------------------------------
When measuring a set of observables, $\{ O \}$, clustering can strongly affect their values by pulling particles into and out of the various regions where $\{O\}$ is measured.  When calculating these effects on the differential cross section, it is useful to separate the calculation into two parts -- the squared matrix element, $\ca{A}$, and the measurement function, $\ca{M}$, which implements the measurement of $\{ O \}$ on the final state.   The measurement function depends on the final state momenta at a given point in phase space, $\Phi$, and the observables being measured. It is integrated against $\ca{A}$ when evaluating the contribution to the cross section:
\be \label{eq:sigmagen}
\frac{d \sigma}{d \{ O \} }
 = \int d\Phi \, \ca{A}(\Phi) \, \ca{M}\big( \{ O \}, \Phi  \big) \,.
\ee
We will be concerned with understanding the general structure of the measurement function for a wide range of  observables for jets defined by a clustering algorithm. This requires translating the iterative recombination procedure of clustering algorithms into all-orders properties of the measurement function -- it is a nontrivial task.  

\begin{figure}[t]
	\begin{center}
	\includegraphics[width=.45\textwidth]{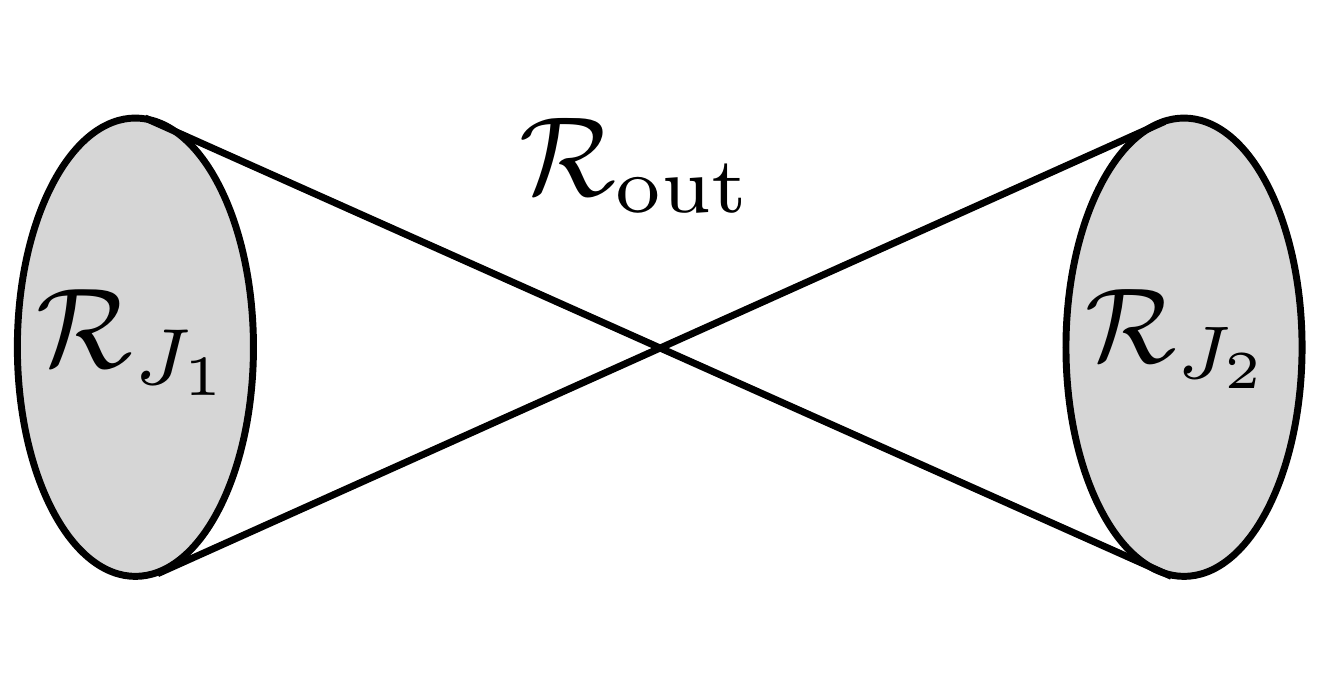} 
	 \end{center} 
	  \vspace{-1em}
{\caption[1]{ Phase space regions $\R_r$ over which observables $O_r$ are measured, contributing to a dijet measurement function. }
\label{fig:Rregions} }
\end{figure}

The measurement function is composed of two main parts: a function of momenta, $f_r(\{k\})$, associated with the measured observable $O_r \in \{ O \}$ and a restriction, $\R_r(\{ k\},\Phi)$, on the region of phase space over which we measure it.  Here $\{k\}$ is a subset of all momenta for a given phase space point,  $\Phi$, that $\R_r$ requires to be in region $r$.
The measurement function can be expressed schematically as 
\be\label{eq:genMeasMom}
\ca{M} \big( \{O\} , \Phi  \big) = \prod_r \, \delta\Big( O_r - \sum_{ \{ k \} }f_r \big(\{k\}\big) \, \R_r \big(\{k\}; \Phi \big) \Big) \,,
\ee
where the product is over the regions of phase space, $r$, which contribute to the cross section.  For example, the constraints that contribute to a dijet cross section are illustrated in \fig{Rregions} and correspond to the two jets, $\R_{J_1}$ and $\R_{J_2}$, and an out-of-jet region, $\R_{\rm out}$, where a different function of momenta, $f_r(\{k\})$, may be measured in each region.
It is often simpler to work with the Fourier transform of the measurement function with respect to the observables:
\begin{align}\label{eq:genMeasPos}
\cM \big( \{x\}, \Phi \big) &= \prod_r \int_{-\infty}^{\infty} {\rm d} O_r \; e^{-ix_r O_r} \, \ca{M} \big( \{ O \}, \Phi  \big) \nn \\
&= \prod_r \exp\left( -ix_r \, \sum_{ \{ k\} }f_r \big(\{k\}\big) \, \R_r \big(\{k\}; \Phi \big)  \right) \,.
\end{align}

An illustrative example is the measurement of the invariant mass of a jet, $m_J^2$; it contributes a factor to $\ca{M}$ of
\be \label{eq:jetmass}
\ca{M} \big(m_J^2, \Phi  \big) = \delta \left( m_J^2 - \sum_{i,j} k_i\cdot k_j  \, \R_J\big(\{k_i,k_j\}; \Phi \big)  \right) \,,
\ee
where $\R_J\big( \{ k_i,k_j \}; \Phi\big)$ gives the phase space constraints for particles $i,j$ to be in the jet $J$.  Expressed explicitly, $\R_J$ is given by a product of theta functions of momenta in $\Phi$ that constrain $k_i$ and $k_j$ to be in the jet.  In general, whether $k_i$ and $k_j$ are in the jet depends on the clustering procedure of the jet algorithm, meaning $\R_J$ depends on all final state momenta $\Phi$.  While the form of \eq{genMeasMom} is compact, it hides correlations between particles, since $\R_J$ depends on momenta not in the jet.  When the measurement function is used in fixed order calculations, the hidden phase space constraints from clustering must be made more explicit. However,  the form of \eq{genMeasMom} makes the perturbative structure of the jet cross section difficult to understand analytically without explicitly carrying out higher order calculations.

The JADE algorithm \cite{Bartel:1986ua} provides an example of why it is important to make manifest the effects of clustering that are hidden in $\R_r$ in \eq{genMeasMom}. Although JADE is a well-defined infrared (IR) safe algorithm that combines particles based on an invariant mass metric, clustering effects at higher orders change the structure of the leading log series ($\as^n \ln^{2n}$) and spoil resummation at this order. This fact is obscured by expressing the action of the algorithm solely through the iterative application of its metric. It required an analytic calculation at two loops \cite{Brown:1990nm} to understand the perturbative structure of jet cross sections defined with JADE. More recently it has been shown that clustering effects for the C/A and $\kt$ algorithms change the structure of the perturbative series at $\ca{O}(\as^2 \ln^2)$ by explicit calculation of the Abelian terms, likely spoiling resummation at this order. Such effects are absent for anti-$\kt$, providing a theoretical preference for this algorithm.

These examples illustrate the importance of having an analytic understanding of the perturbative structure of jet cross sections. However higher order analytic calculations are difficult and can be made prohibitively complex by clustering phase space constraints. Since the structure of the measurement function, $\ca{M}$, determines clustering effects, we focus on expressing $\ca{M}$ in a form that makes these effects manifest.  This will provide insight into the perturbative structure of jet cross sections without having to carry out explicit calculations.

%------------------------------------------------------------------------------------------------------------------------------------------------
\subsection{Reformulating the Constraints from Clustering}
%------------------------------------------------------------------------------------------------------------------------------------------------
Running a clustering jet algorithm divides up phase space into mutually exclusive regions that cover all of phase space.  For an $N$-jet observable theses regions may correspond to each jet, $\R_{J_1},\ldots,\R_{J_N}$, and an out-of-jet region $\R_{\rm out}$. The jet algorithm does this by separating final state particles into groups that characterize the outcomes of the algorithm.  For example, for three particles with momentum $k_1$, $k_2$, and $k_3$, the possible outcomes, $\ca{K}_{\ell}(\Phi)$, of an algorithm are
\begin{itemize}
\item $k_1$, $k_2$, and $k_3$ all cluster together: $\ca{K}_{3}$   \vspace{-0.5em}
\item $k_1$ and $k_2$ cluster, but $k_3$ does not (+ permutations): $\ca{K}_{12}, \ca{K}_{13}$, and $\ca{K}_{23}$   \vspace{-0.5em} 
\item None of the particles cluster: $\ca{K}_{1}$ 
\end{itemize}
Each outcome $\ca{K}_{\ell}(\Phi)$ represents a set of phase space constraints requiring the particles to cluster into a set of groups.  Summing over all possible outcomes, $\ell$, of a clustering algorithm at a given number of final state particles gives a unitarity relation:
\be \label{eq:genKunitarity}
1=\sum _{\ell} \ca{K}_{\ell} (\Phi) \,.
\ee
Since the $\ca{K}_{\ell}$ makes the phase space constraints for a given outcome of the clustering algorithm explicit, $\R_r$ is simplified.  $\R_r \big(\{k\}; \Phi\big)$ constrains $\{k\}$ to be in region $r$ where the observable $O_r$ is measured, but it no longer must implement the constraints from clustering, which appear in the $\ca{K}_{\ell}$.  This means $\R_r$ depends only on the momenta in the region $r$:
\be
\R_r \big(\{k\}; \Phi\big) \quad \to \quad \R_r \big(\{k\}\big) \,.
\ee
To write the measurement function, we simply weight \eq{genKunitarity} by \eq{genMeasMom}: 
\be \label{eq:genMeasK}
\ca{M} \big(\{O\},  \Phi  \big) = \sum_{\ell} \ca{K}_{\ell} (\Phi) \prod_r \, \delta\Big( O_r - \sum_{ \{ k\} }f_r \big(\{k\}\big) \, \R_r \big(\{k\} \big) \Big) \,.
\ee

While it is simple to enumerate the outcomes of clustering, each $\ca{K}_{\ell}$ contains a complicated set of phase space constraints; as such, in order to explore this structure we need to know where the IR divergences arise, since they can contribute large logarithms in the perturbative series of the cross section.  This depends on both the squared matrix element, $\ca{A}$, and the measurement, $\ca{M}$. Although the structure of $\ca{A}$ is complicated at higher orders, we will see that examining the structure of $\ca{M}$ alone can provide insight. The differential cross section is sensitive to the IR regions of phase space determined by $\R_r$ in \eq{genMeasPos}. 

\begin{figure}[t]{
	\begin{center}
	\includegraphics[width=1\textwidth]{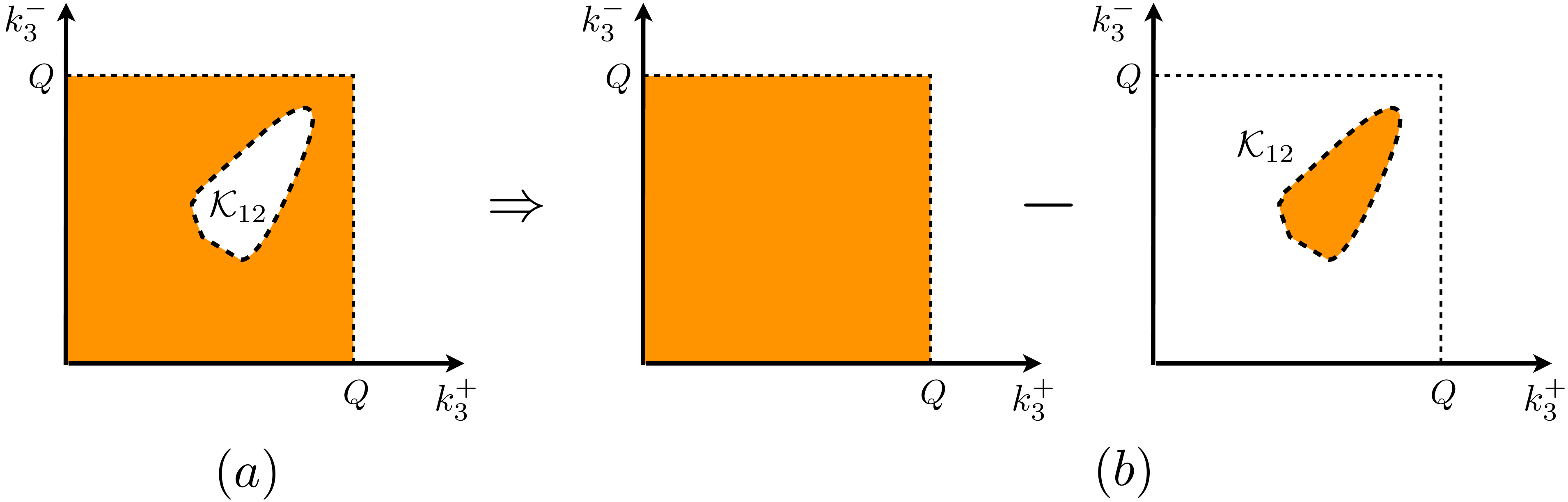} 
	  \end{center} 
	  \vspace{-1em}
{ \caption[1]{ Phase space for particle $k_3$ for fixed $k_1, k_2$. In the region $\ca{K}_{12}$, $k_3$ would be clustered with $k_1$ and $k_2$. The exclusive constraint $\theta(k_3 \not\in \ca{K}_{12})$ is shown in (a) by the orange region and the inclusive constraint $1 - \theta(k_3 \in \ca{K}_{12}) $ is shown in (b) by the orange region.}
  \label{fig:q3R12}} }
\end{figure}

In this paper we provide a framework to use unitarity relations of the type \eq{genKunitarity} to express the outcome of the clustering algorithm and isolate the most divergent regions of phase space, which contribute the most logarithms in the perturbative expansion. As we shall see, this is done by using unitarity iteratively to rewrite \textit{exclusive} constraints as \textit{inclusive} ones. For example, the case where $k_1$ and $k_2$ cluster but $k_3$ does not contains a phase space constraint excluding $k_3$ from some region of phase space; it cannot be close enough to $k_1$ or $k_2$ to fall in one of the other outcomes. This is shown in \fig{q3R12}(a), where for a given $k_1$ and $k_2$ there is a region $\ca{K}_{12}$ that $k_3$ can not occupy in order to avoid being clustered\footnote{We decompose the four-momentum as
\begin{align}
k^\mu =  n\cdot k \frac{ \bn^\mu}{2}+ \bn \cdot k \frac{n^\mu}{2}+k_\perp^\mu \, , \nn
\end{align} 
and define $k^+ \equiv n\cdot k$ and $k^- \equiv \bn \cdot k$. Here $n$ and $\bn$ are light-cone vectors, $n^{\mu} = (1,\hat{n})$ and $\bn^{\mu} = (1,-\hat{n})$, with $\hat{n}^2 =1$.}. Divergences in phase space arise as $k_3^+,k_3^- \to 0$ along the axes and the associated logarithms will depend on $k_1$ and $k_2$ in a nontrivial way through the boundary of $\ca{K}_{12}$.  A more useful description of clustering would rewrite this in terms of an inclusive phase space constraint shown in \fig{q3R12}(b), where the exclusive and inclusive ways to write this constraint would be
\begin{align}
\text{exclusive : } & \theta(k_3 \not\in \ca{K}_{12}) \,, \nn \\
\text{inclusive : } & 1 - \theta(k_3 \in \ca{K}_{12}) \,.
\end{align}
The function $\theta(x)$ is the usual step function, giving 1 if $x>0$ and 0 otherwise. The inclusive description expresses the divergences in $k_3$ in a way that is independent of $k_1$ and $k_2$ and subtracts a finite contribution from the region $\ca{K}_{12}$. As a result, integrating over the momenta of particle $k_3$ with the inclusive constraints simplifies the divergent structure \cite{Bauer:2011hj,Jouttenus:2011wh}. 
Rewriting the exclusive constraint as an inclusive one uses a (simple) unitarity relation describing the phase space constraints between particles. One can iteratively apply unitarity relations of this kind to rewrite the outcome of clustering only in terms of inclusive phase space constraints. 

%-------------------------------------------------------------------------------------------------------------------
\subsection{An Application: Soft Clustering and Jet Shapes}
%-------------------------------------------------------------------------------------------------------------------
Energetic particles have the ability to change the number, energy, and direction of jets.  Often, though, energetic particles are confined to small regions of phase space and only cluster within jets.  This occurs for jet observables where the measurement of an observable $O$ constrains the collinear radiation to be tightly collimated within its own jet, compared to the jet size; e.g. measuring the jet mass, $m_J$, when  $m_J /E_J \ll  R$ forces this to be true.  This is satisfied for a range of jet shapes and substructure observables \cite{Ellis:2009wj,Ellis:2010rwa,Thaler:2010tr,Jankowiak:2011qa,Thaler:2011gf} in the soft/collinear regime.  In this case the boundaries of the jets are determined by clustering amongst \emph{soft} particles.

In this regime the function of momenta that we measure, $f_r( \{k \} )$ in \eq{genMeasPos}, simplifies to be a function of momentum of a \textit{single} particle $f_r(k_i)$ for a large class of observables.  For the example of jet mass given in \eq{jetmass}, the expression greatly simplifies to give
\be \label{eq:Msoft}
\sum_{i,j} k_i\cdot k_j  \, \R_J\big(\{k_i,k_j\}\big) \quad \to \quad \sum_i p_J\cdot k_i \, \R_J \big(k_i \big) \, ,
\ee
where $p_J$ is the sum of the collinear momentum and is conserved for each jet direction in the soft/collinear regime. The measurement function still contains correlations between particles through the outcomes of the algorithm, given by $\ca{K}_{\ell}$:
\be\label{eq:fsingle}
\cM \big(\{x\},  \Phi \big)= \sum_{\ell} \ca{K}_{\ell} (\Phi) \prod_r \exp\left( -i \, \sum_{i} x_r \,  f_r \big(k_i \big) \, \R_r \big(k_i \big)  \right) \,.
\ee

Soft clustering for the anti-$\kt$ algorithm is particularly simple: jets in the soft/collinear regime are circular with jet size $R$ around the collinear direction up to power suppressed corrections \cite{Cacciari:2008gp,Ellis:2010rwa,Walsh:2011fz}. In this case the sum over outcomes in \eq{fsingle} simplifies to a single outcome: none of the particles cluster, and the constraint is trivial, $\ca{K}(\Phi)=1$. This removes all correlations between different particles in the final state from the measurement function. As a result the measurement simplifies greatly:
\begin{align}\label{eq:Mfact}
\cM_{\akt} \big(\{x\},  \Phi  \big)
= \prod_{i} \cM^{(1)} \big(\{x\}, k_i \big)
\,.
\end{align}
where
\be\label{eq:M1akt}
\cM^{(1)} \big(\{x\}, k  \big) = \exp \Big(-i \, \sum_{r} x_r \, f_r (k ) \, \R_r (k )  \Big)  \,.
\ee
It is well known that the measurement function also takes this simple form for a large class of event shapes in the soft/collinear limit, such as thrust, heavy jet mass, angularities, and N-jettiness \cite{Berger:2003iw,Bauer:2008dt,Hornig:2009vb,Stewart:2010tn}, which makes such shapes theoretically appealing. Notice that jet rates do not satisfy the form in \eq{Mfact} except for fixed cone algorithms, which have known split/merge problems \cite{Ellis:2007ib,Salam:2009jx}.

\begin{figure}[t]{
	\begin{center}
	\includegraphics[width=.75\textwidth]{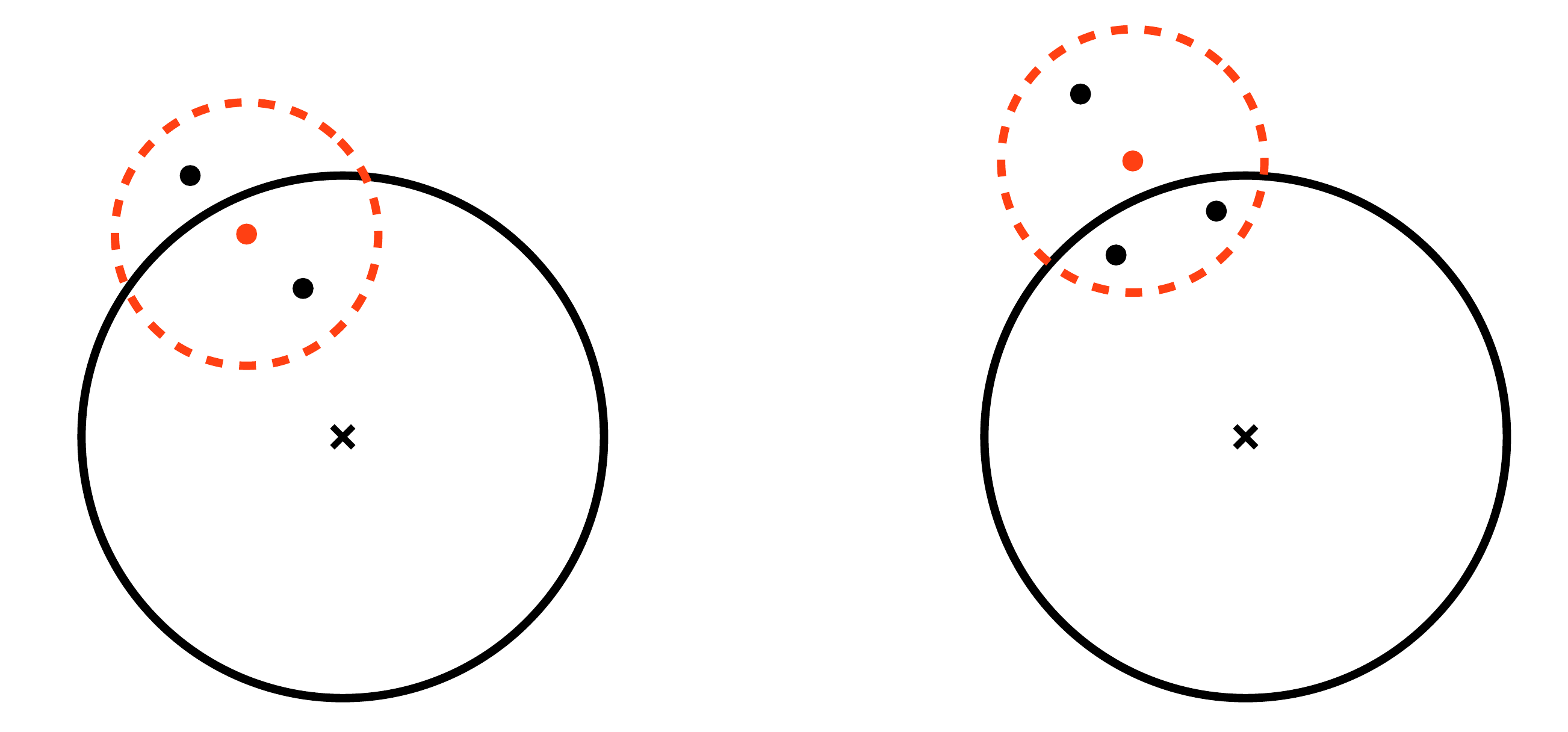} 
	  \end{center} 
	  \vspace{-1em}
{ \caption[1]{Configurations of two (left) and three (right) particle boundary clustering that changes the jet boundary.  The final state particles shown in black cluster, and the red particle is the result.  In the two particle case the particle outside the jet is pulled in by clustering, and in the three particle case the particles in the jet are pulled out.}
  \label{fig:BoundaryClusEx}} }
\end{figure}

Algorithms such as C/A or $\kt$ modify the phase space regions from the simple cones of the anti-$\kt$ algorithm,  due to  clustering amongst soft gluons.   This occurs when a soft particle is pulled in or out of the jet, and we refer to this as
\begin{align}
\textit{boundary clustering}\, \text{ : }& \text{clustering of soft particles across the jet boundary of radius $R$} \nn \\
&\text{around the jet direction.} \nn
\end{align}
Example configurations for 2 and 3 particles are shown in \fig{BoundaryClusEx}.  
Since boundary clustering modifies the region over which the measurement is made and is an effect that intrinsically depends on the momenta of all particles that are boundary clustered, the phase space constraints for the outcome, $\ca{K}_{\ell}(\Phi)$, of the C/A or $\kt$ measurement functions will depend on multiple momenta in the final state. This introduces correlations into the measurement function, which in this case is of the form \eq{fsingle}. A useful way to organize the correlations introduced by $\ca{K}_{\ell}(\Phi)$ is through a correction to the measurement function of the form  
\be\label{eq:Mcorr}
\cM^{(n)} \big( \{x\}, \Phi \big) = \prod_{i = 1 }^{n} \cM^{(1)} \big(\{ x\},  k_i \big) + \cM_{\rm corr.}^{(n)} \big( \{x\},  \Phi \big) \,,
\ee
where $\cM^{(1)}$ is given in \eq{M1akt} and $n$ is the number of final state particles. This is obtained from \eq{fsingle} by a trivial application of the unitarity relation $\ca{K}_{\ell}(\Phi) \equiv 1-\overline{\ca{K}}_\ell(\Phi)$ for each outcome $\ell$, where this relation defines $\overline{\ca{K}}_\ell(\Phi)$. 

Clustering for algorithms other than anti-$\kt$ affect the logarithmic terms arising from soft divergences. In the soft and collinear regimes, boundary clustering can lead to large logarithms in the cross section, starting at $\ca{O}(\as^2)$.  This was shown in \cite{Banfi:2005gj,Delenda:2006nf} and further explored in \cite{KhelifaKerfa:2011zu,Kelley:2012kj}, and the original authors termed the effect \textit{clustering logs}.  In \cite{Kelley:2012kj} it was shown that clustering logs have the same properties as non-global logarithms (NGLs) \cite{Dasgupta:2001sh,Dasgupta:2002bw} in the soft/collinear limit and that they contribute at least at next-to-leading logarithm\footnote{We count in the exponent of the distribution, where N${}^k$LL terms are of order $\as^n \ln^{n-k+1}$.} (NLL), but could in principle contribute at LL.  It was also argued that new clustering effects likely arise at each order in $\as$ at NLL and beyond, spoiling resummation.   

Using unitarity relations of the form \eq{genKunitarity}, we can express the measurement function to isolate the divergences from boundary clustering. These corrections can be parameterized in terms of the number of particles involved in clustering.  In particular we show 
\be\label{eq:corrStruct}
\cM_{\rm corr.}^{(n)} \big( \{ x\}, \Phi \big) = \sum_{k=0}^{n-2} \frac{n!}{(n-k)! \, k!} \, \Delta \cM^{(n-k)} \big[\cM^{(1)}\big]^k \,,
\ee
where $\Delta \cM^{(k)}$ is the contribution where $k$ particles cluster with at least one other particle and first arises at $\ord{\as^k}$. This result applies not only to boundary clustering, but clustering in general, away from soft-collinear regime. While clustering effects can a priori contribute at LL, the form of the measurement function in \eq{corrStruct} allows us to show that clustering logarithms arise at NLL and to determine to all orders the form of the Abelian soft function and show that exponentiation of these terms does not occur.

%-----------------------------------------------------------------------------------------------------------
\subsection{Outline}
%-----------------------------------------------------------------------------------------------------------
In this paper, we construct a framework to describe the phase space constraints from clustering and the effect on observables.  The formalism we use is very general, applying to a wide class of clustering algorithms and measurements.  It can be applied to a specific type of clustering, such as boundary clustering, or to clustering in general.  The formalism to describe clustering is presented in two main parts.

First, in \sec{unitarity}, we define a set of functions that schematically describe the phase space constraints from clustering.  They are functions of clustered groups of particles, which are the possible outcomes of the jet algorithm. The physical interpretation of these functions allows us to construct unitarity relations that transform exclusive constraints into inclusive ones.  The result is a prescription to systematically determine the phase space constraints from clustering in terms of inclusive constraints.  Our goal is not to write clustering constraints explicitly in terms of the momenta of individual particles, but instead to express the constraints in a form that highlights the analytic properties of the outcome of a generic clustering algorithm.  We do not deal directly with the individual clustering steps of the algorithm, but in most cases our results can be applied to describe each step.

Second, in \sec{measurement} we apply this framework to the measurement function.  For any measurement function of the form \eq{genMeasK} where the observables depend on the momenta of a single clustered group, we use the results of \sec{unitarity} to find the corrections due to clustering. We show that these have the form \eq{corrStruct}. This form can be exploited to determine the organization of logarithms arising from clustering.  We also comment on the infrared safety of clustering algorithms in \sec{IRdiv}; we can use this framework to see precisely how clustering effects among soft particles are IR safe.

In \sec{AbelianAllOrders}, we apply the formalism of the measurement function to study the logarithmic structure of the soft function arising from clustering.  We focus on the Abelian terms, where the matrix elements factorize and it is straightforward to translate the measurement function into contributions to the cross section.  We use the form of the measurement function along with constraints from renormalization to determine the all-orders structure of the soft function, allowing us to show that clustering NGLs contribute at most at NLL at all orders in $\as$.  Finally, in \sec{conclusions}, we summarize our results.

%%%%%%%%%%%%%%%%%%%%%%%%%%%%%%%%%%%%%%%%%%%%%%%%%%%
\section{Phase Space Constraints from Clustering Algorithms}
\label{sec:unitarity}
%%%%%%%%%%%%%%%%%%%%%%%%%%%%%%%%%%%%%%%%%%%%%%%%%%%

In this section and the next, ``clustering'' can refer to general clustering or to a specific kind, such as boundary clustering.  If this framework is used to describe boundary clustering, which is relevant for observables that constrain the collinear radiation in the jet to be collimated compared to the jet size (e.g. $m_J/E_J \ll R$), then the phase space constraints will ignore clustering that does not change the jet boundary.  We will also treat the particles as indistinguishable; this is not necessary, but simplifies the discussion.

As discussed in the introduction, the deterministic nature of jet algorithms allows the action of the jet algorithm to be characterized in terms of a set of mutually exclusive outcomes, $\ca{K}_{\ell}$, as in \eq{genKunitarity}.  When we run the jet algorithm, sets of particles will cluster, and we use these sets to characterize the outcomes of the algorithm.  For example, if there are 5 particles in the final state labeled $q_1, \ldots, q_5$, then a possible outcome from a jet algorithm $\ca{J}$ is
\be \label{eq:clusteringexample}
\ca{J}: \{q_1, q_2, q_3, q_4, q_5\} \quad \longrightarrow \quad \{ \{ q_1, q_2 \}, \{q_3, q_4\}, \{q_5\}\} \,.
\ee
In this case particles 1 and 2 have been clustered, as have particles 3 and 4, and particle 5 does not cluster.

This example shows that the action of a jet algorithm is to partition the final state particles into groups. We refer to partitions like the one above as labeled partitions.  There is a one-to-one association between labeled partitions and outcomes of the algorithm.  However, if the particles are indistinguishable, there are many labeled partitions that describe the same basic outcome.  In the case of \eq{clusteringexample}, the important characteristic is that particles were clustered in groups of 2, 2, and 1; a distinct clustering that is equivalent to this outcome is $\{\{q_1, q_4\}, \{q_2, q_5\}, \{q_3\}\}$.  Therefore, we can simplify the labeled partitions and describe them in terms of the number of particles clustered in each group.  We will refer to these simply as partitions.  The labeled partition in \eq{clusteringexample} maps to the partition $\{2,2,1\}$.  For 5 particles, the possible partitions are
\be
\{5\},\, \{4,1\},\, \{3,2\},\, \{3,1,1\},\, \{2,2,1\},\, \{2,1,1,1\},\, \{1,1,1,1,1\}\,.
\ee

In general, we will denote partitions by $\ca{P} = \{ p_1 , \ldots, p_k \}$, where the $p_i$ are natural numbers.  The number of labeled partitions that are equivalent to the partition $\ca{P}$ is
\be \label{eq:NPdef}
N(\ca{P}) = \frac{n!}{p_1! \cdots p_k!} \frac{1}{d_1! \cdots d_n!} \,,
\ee
where $n$ is the total number of particles, and $d_i$ is the number of elements of $\ca{P}$ equal to $i$.  The first factor in $N(\ca{P})$ is the usual multinomial coefficient, and the second factor accounts for the degeneracy in dividing particles among equal sized elements in the partition.  

%%%%%%%%%%%%%%%%%%%%%%%%%%%%%%%%%%%%%%%%%%%%%%%%%%%
\subsection{Unitarity Relations and Phase Space Constraints}
\label{ssec:PSconstraints}
%%%%%%%%%%%%%%%%%%%%%%%%%%%%%%%%%%%%%%%%%%%%%%%%%%%

We will now discuss how we can more specifically determine the phase space constraints $\ca{K}_{\ell}$, defined in \sec{intro},  for each outcome, $\ell$, of the jet algorithm (which we associate with a partition).  In words, a partition represents the phase space constraints:
\begin{align} \label{eq:Pdef}
\{p_1, \ldots, p_k\} \, &: \, \text{$p_1$ particles cluster, $p_2$ particles cluster, $\ldots$, $p_k$ particles cluster,} \nn \\
& \qquad \text{and these groups of particles do not cluster with each other.}
\end{align}
Therefore, we must define two basic functions that act on the elements of a partition: one requiring a set of particles to cluster and one requiring different groups of particles not to cluster.  These can be defined in terms of objects in the partition:
\be \label{eq:cdef}
c(p)\, : \, p \text{ particles are clustered,}
\ee
and
\be \label{eq:sdef}
s\cset{p_1, \ldots, p_k} \, : \, \text{the clusters of $p_1,\ldots,p_k$ particles do not cluster.}\vspace{0.5em}
\ee
For completeness, we define $s(\{p\}) \equiv 1$ and $c(1) \equiv 1$ (i.e., no constraint).  The function $s\cset{p_1, \ldots, p_k}$ defines an \textit{exclusive} constraint; for example, it excludes the set of particles $p_k$ from some region of phase space so that they do not combine with the sets of particles $p_1, \ldots, p_{k-1}$.  For a given partition $\ca{P}$, the phase space constraints we assign to it are
\be
\bigg[ \prod_{p\in\ca{P}} c(p) \bigg] s(\ca{P}) \,.
\ee
If we sum over all possible outcomes of the jet algorithm, we obtain a unitarity relation for phase space:
\be \label{eq:clusteringunitarity}
1 = \sum_{\ca{P}} N(\ca{P}) \bigg[ \prod_{p\in\ca{P}} c(p) \bigg] s(\ca{P}) \,,
\ee
where the sum is over partitions.  Each term in the sum corresponds to an outcome $\ca{K}_{\ell}$ of the algorithm, so that \eq{clusteringunitarity} is a more specific version of \eq{genKunitarity}.

\begin{figure}[t]{
	\begin{center}
	\includegraphics[width=.9\textwidth]{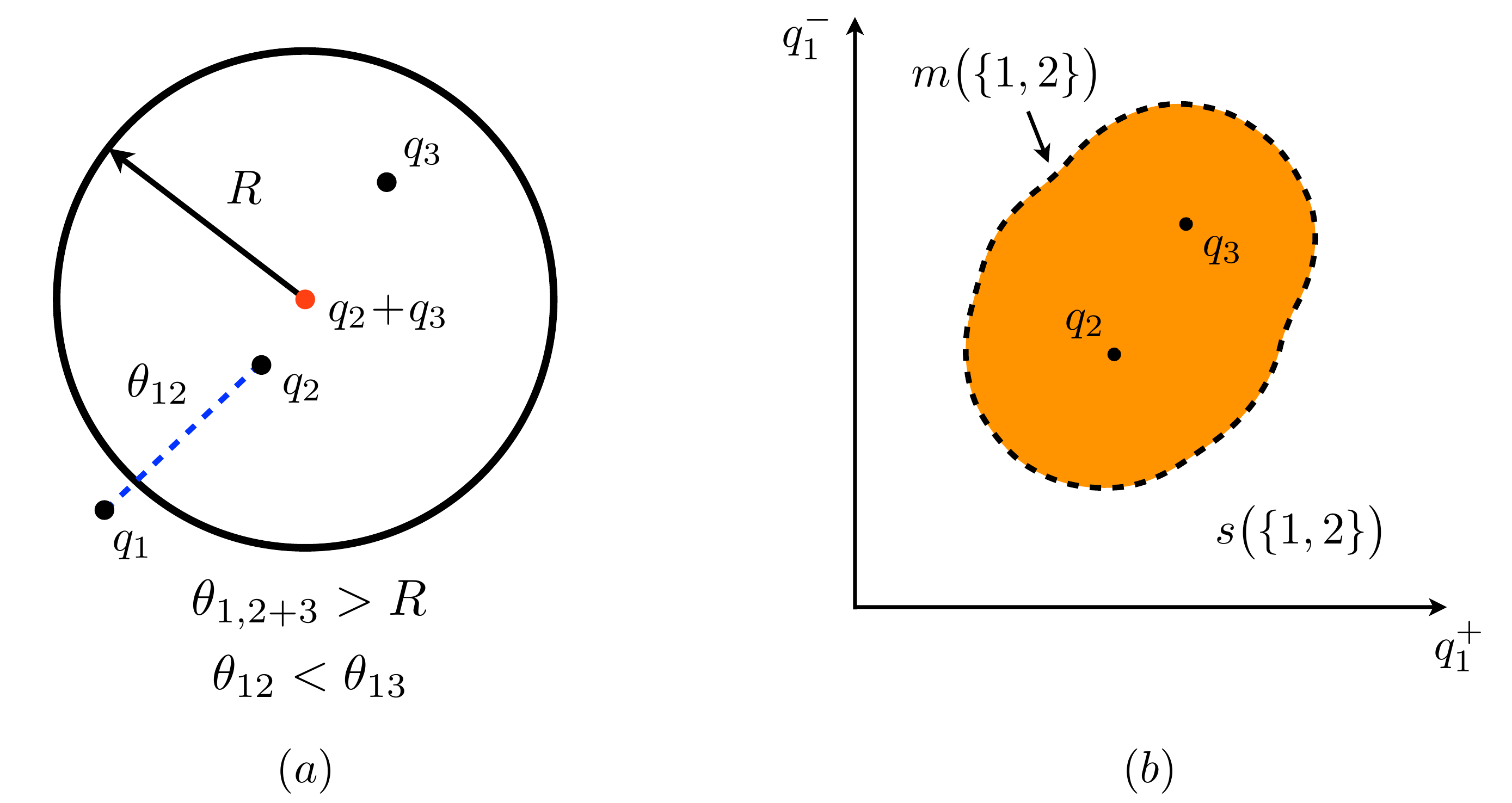} 
	  \end{center} 
	  \vspace{-1.5em}
{ \caption[1]{On the left, a configuration of 3 particles for the C/A algorithm that contributes to $c(2) \, m\cset{1,2}$ but not to $c(3)$. ÊIn this case particles $q_2$ and $q_3$ cluster first to form the merged pair $q_2 + q_3$, shown in red. ÊParticle $q_1$ is too far away to cluster with $q_2 + q_3$, as it lies outside the solid circle of radius $R$ centered at $q_2 + q_3$. ÊTherefore the 3 particles do not cluster, and so do not contribute to $c(3)$. ÊOn the right, a schematic picture of the region of phase space for particle $q_1$ defined by $m\cset{1,2}$, which is the complement of $s\cset{1,2}$.}
 Ê\label{fig:m21config}} }
\end{figure}

Some basic examples are useful.  For two particles, the partitions are $\{1,1\}$ and $\{2\}$, and the unitarity relation is
\be
1 = c(2) + s\cset{1,1} \,.
\ee
This relation defines $s\cset{1,1}$.  For three particles, the partitions are $\{1,1,1\}$, $\{1,2\}$, and $\{3\}$, and the unitarity relation is
\be
1 = c(3) + 3\,c(2) s\cset{1,2} + s\cset{1,1,1} \,.
\ee
Consider the function $s\cset{1,2}$.  Let us label the single particle $q_1$ and the pair of particles that cluster $q_2$ and $q_3$.  The function $s\cset{1,2}$ allows $q_1$ to go anywhere except for near $q_2$ and $q_3$.  However, we can use a unitarity constraint to rewrite $s\cset{1,2}$ in terms of an \textit{inclusive} constraint:
\begin{align}\label{eq:s21}
s\cset{1,2} &= (\text{$q_1$ goes anywhere}) - (\text{$q_1$ would cluster with $q_2, q_3$, or $q_2+q_3$})\,. \nn \\
&\equiv 1 - m\cset{1,2} \,.
\end{align}
This relation defines the \textit{merging} function $m\cset{1,2}$, which requires $q_1$ to be near the clustered pair.  However, this function does not require that all three particles would cluster when running the jet algorithm:
\be
c(2) \,m\cset{1,2} \neq c(3) \,.
\ee
For example, for fixed momenta of $q_2$ and $q_3$, the single particle $q_1$ could be close enough to $q_2$ but sufficiently far from $q_3$ so that all three particles would not cluster.  In \fig{m21config}, we show a configuration that is part of $c(2) m\cset{1,2}$ but not part of $c(3)$, as well as a schematic region of phase space for $m\cset{1,2}$. Only in the case of two particles does $m\cset{1,1} =c(2)$.
In general, merging requires that at least one particle from each group would cluster with at least one particle in another group, such that all the groups are connected.  The general merging function is
\be \label{eq:mdef}
m\cset{p_1, \ldots, p_k} \, : \, \text{the clusters of $p_1, \ldots, p_k$ particles merge.}
\ee

Our general goal will be to rewrite the exclusive $s(\ca{P})$ functions in terms of the inclusive clustering and merging functions that require particles to be grouped together.  For each function $s(\ca{P})$, we can define a unitarity relation to rewrite $s(\ca{P})$ in terms of clustering and merging functions.  Note that rewriting $s(\ca{P})$ does not change the outcome of the jet algorithm, but instead rewrites the phase space constraints from clustering in the measurement function in a way that makes the perturbative structure of the cross section clearer.

\begin{figure}[th!]{
\begin{center}
 \includegraphics[width=\textwidth]{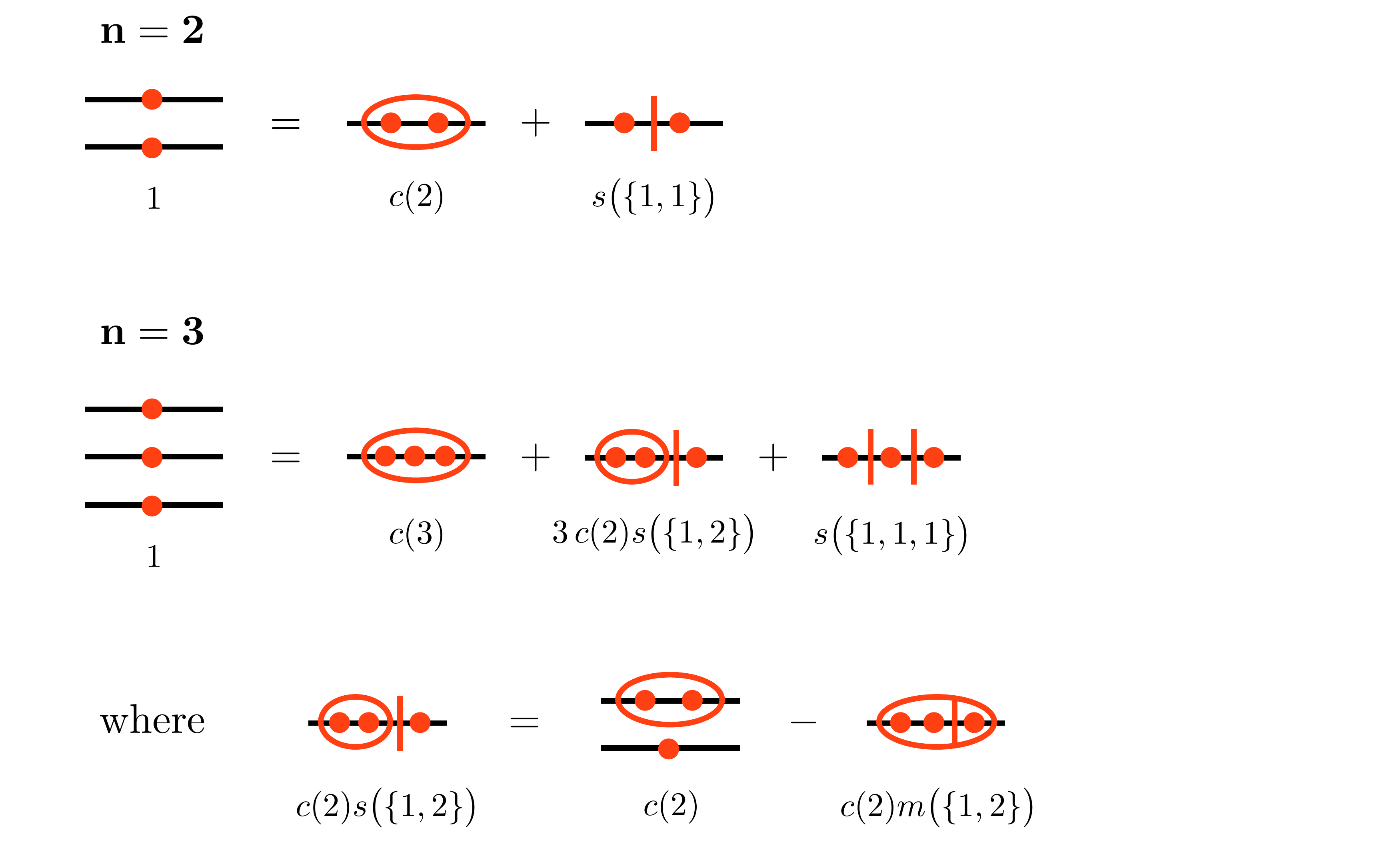} 
\end{center}
{ \caption[1]{Clustering diagrams for the cases $n=2$ and $n=3$.  Particles are represented by dots, and clustered particles are circled.  Particles that do not cluster have a vertical bar between them.  Merged groups of clustered particles are circled, with a vertical  bar between each group.  The different terms in the unitarity relations are shown. We also show how to express the $s(\ca{P})$ functions requiring particles to not cluster in terms of $c(p)$ and $m(\ca{P})$ functions that require the particles to be clustered/merged.}
\label{fig:Cases2and3}} }
\end{figure}

We introduce some diagrammatic notation which allows us to express these unitarity relations in a simple way.  These diagrams are helpful in thinking about higher order particle configurations and are useful in determining unitarity relations.  Instead of clustering in 3-space, it is sufficient to think of clustering along a line.  Each particle will be depicted as a dot, and if particles are clustered then we circle them.  If the particles do not cluster, we put a vertical bar between them.  If particles are not on the same line, then their phase space constraints do not depend on each other.
Finally, when we require groups of particles to merge, we circle those groups and place a vertical bar between the groups.  In \fig{Cases2and3}, we give the diagrams for the 2 and 3 particle cases.
%

%%%%%%%%%%%%%%%%%%%%%%%%%%%%%%%%%%%%%%%%%%%%%%%%%%%
\subsection{Clustering Maps}
\label{ssec:clusteringmaps}
%%%%%%%%%%%%%%%%%%%%%%%%%%%%%%%%%%%%%%%%%%%%%%%%%%%

When we rewrite the phase space constraints given by $s(\ca{P})$ in \eq{clusteringunitarity}, we can represent the result as a map between partitions.  Consider the unitarity relation that defines $s\cset{1,1,1}$ in \fig{Cases2and3}:
\begin{align} \label{eq:s111rel}
s\cset{1,1,1} &= 1 - c(3) - 3\,c(2)s\cset{1,2} \nn \\
&= 1 - c(3) - 3\,c(2) + 3\,c(2)m\cset{1,2} \,.
\end{align}
Each term represents a different phase space constraint on the three particles, and correspondingly a map to a different partition.
\begin{equation}
\begin{tabular}{c c l}
$1                   $ &:& $ \quad \{1,1,1\} \quad \to \quad \{1,1,1\}                $ \\
$-3\,c(2)            $ &:& $ \quad \{1,1,1\} \quad \to \quad \{1,2\}                  $ \\
$3\,c(2) m\cset{1,2} $ &:& $ \quad \{1,1,1\} \quad \to \quad \{1,2\} \quad \to \quad \{3\}  $ \\
$-c(3)               $ &:& $ \quad \{1,1,1\} \quad \to \quad \{3\}                    $ \\
\end{tabular}
\end{equation}

Each of these maps represent a coarsening of the partition $\{1,1,1\}$.  A partition $\ca{P}_1$ is \textit{coarser} than a partition $\ca{P}_2$ if $\ca{P}_1$ is made by combining elements of $\ca{P}_2$ ($\ca{P}_2$ is \textit{finer} than $\ca{P}_1$).  These maps can be used to determine all ways to rewrite a given $s(\ca{P})$.  For a given partition $\ca{P}$, we write all maps to coarser partitions (or itself), and each step in the map is associated with a factor that either merges or clusters particles.  For instance, in the map $\{1,1,1\} \to \{1,2\} \to \{3\}$, the associated factors are
\begin{equation}
\begin{tabular}{r c c c c c}
map : &$\{1,1,1\} $ & $ \quad \longrightarrow \quad $ & $ \{1,2\} $ & $\quad \longrightarrow \quad$ & $\{3\} \, $ \\ 
contribution : & & $-3 c(2) $ & & $-m\cset{1,2}$ &
\end{tabular}
\end{equation}
In the first step, we cluster two single particles, and there are 3 ways to choose a pair of particles.  In the second step, we merge the single particle with the clustered pair.  The minus signs arise because we are using a unitarity relation, as in \eq{s111rel}.  Note that whenever a map involves only grouping single particles, e.g. $\{1,1,1\} \to \{1,2\}$,  a clustering factor is generated, and whenever it involves a group with multiple particles, e.g. $\{1,2\} \to \{3\}$, a merging factor is generated.

We can think of a map, $\phi$, as a sequence that successively coarsens the initial partition:
\be \label{eq:phimap}
\phi(\ca{P}_1,\ca{P}_N) = \{\ca{P}_1, \ldots, \ca{P}_N\} \,, \;\text{ with $\ca{P}_{i+1}$ coarser than $\ca{P}_i$\,.}
\ee
Each step along the sequence is associated with a clustering or merging factor. We define the contribution from each step in the map to not include the functions $s(\ca{P}_i)$ appearing in the full phase space constraints.  To formalize this, we will define a few special partitions:
\begin{align} \label{eq:Pmapdefs}
\ca{P}^0 \, &: \, \text{the base partition $\{1,\ldots,1\}$\,.} \\
\ca{P}^{\de} \, &: \, \text{the partition that gives the outcome of the algorithm} \nn \\
\ca{P}^{f} \, &: \, \text{the final partition in a map $\phi$\,.} \nn
\end{align}
In \eq{clusteringunitarity}, we obtain a factor of $s(\ca{P}^{\de})$ that we want to rewrite in terms of inclusive $c$ and $m$ functions.  The sum over all possible maps $\phi(\ca{P}^{\de}, \ca{P}^f)$, including summing over all possible $\ca{P}^f$,  gives the unitarity relation that lets us rewrite $s(\ca{P}^{\de})$.  By definition, $\ca{P}^{f}$ must be coarser than or equal to $\ca{P}^{\de}$.

\begin{figure}[th!]{
\begin{center}
 \includegraphics[width=\textwidth]{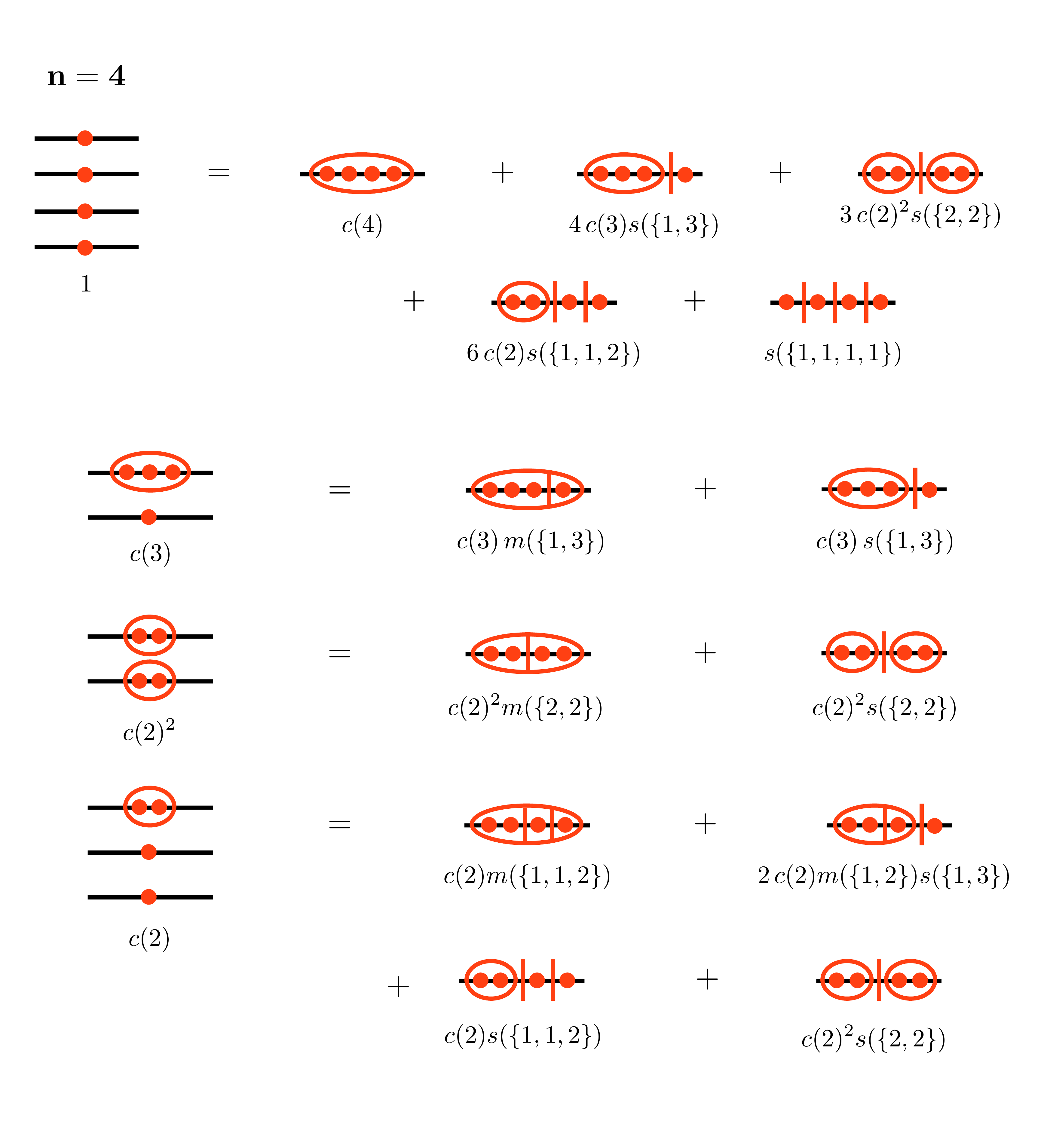} 
\end{center}
{ \caption[1]{Clustering diagrams for the case $n=4$.  The meaning of the diagrams are the same as in \fig{Cases2and3}.  These can be used to derive the unitarity relations for 4 particles, such as $s\cset {1,1,1,1}$.}
\label{fig:Case4}} }
\end{figure}

As an example, for 4 particles we can use these maps to determine the unitarity relation for $s\cset{1,1,1,1} = s(\ca{P}^0)$.  These maps and the associated factors are
\begin{equation}\label{eq:4MeasClusMaps}
\begin{tabular}{c c c c c c c c}
\vspace{1mm}
$\boldsymbol{\phi:}$      & $ \boldsymbol{ \ca{P}^{\de} }$ & & $\boldsymbol {\longrightarrow}$ & & $\boldsymbol{ \ca{P}^{f}  }     $ &\bf{:}&  \bf{contribution                 }     \\
 \vspace{-3mm}\\
\vspace{1mm} & $\ca{P}^0 $ & & $\longrightarrow$ & & $\{1,1,1,1\}         $ &:&  $    1                  $ \\
\vspace{1mm} & $\ca{P}^0 $ & & $\longrightarrow$ & & $\{1,1,2\}           $ &:&  $ -6\,c(2)              $ \\
\vspace{1mm} & $\ca{P}^0 $ & & $\longrightarrow$ & & $\{1,3\}           $ &:&  $ -4\,c(3)              $ \\
\vspace{1mm} & $\ca{P}^0 $ & $\to$ & $\{1,1,2\}$ & $\to$ & $\{1,3\} $ &:& $ [-6\,c(2)][-2\,m\cset{1,2}] $ \\
\vspace{1mm} & $\ca{P}^0 $ & & $\longrightarrow$ & & $\{2,2\}           $ &:&  $ -3\,c(2)^2              $ \\
\vspace{1mm} & $\ca{P}^0 $ & $\to$ & $\{1,1,2\}$ & $\to$ & $\{2,2\} $ &:&  $ [-6\,c(2)][-c(2)] $ \\
\vspace{1mm} & $\ca{P}^0 $ & & $\longrightarrow$ & & $\{4\} $ &:&  $ -c(4) $ \\
\vspace{1mm} & $\ca{P}^0 $ & $\to$ & $\{1,3\}$ & $\to$ & $\{4\}             $ &:&  $ [-4\,c(3)][-m\cset{1,3}]         $ \\
\vspace{1mm} & $\ca{P}^0 $ & $\to$ & $\{2,2\}$ & $\to$ & $\{4\}             $ &:&  $ [-3\,c(2)^2][-m\cset{2,2}]         $ \\
\vspace{1mm} & $\ca{P}^0 $ & $\to$ & $\{1,1,2\}$ & $\to$ & $\{4\}             $ &:&  $ [-6\,c(2)][-m\cset{1,1,2}]         $ \\
\vspace{1mm} & $\ca{P}^0 $ & $\to$ & $\{1,1,2\} \to \{1,3\}$ & $\to$ & $\{4\} $ &:&  $ [-6\,c(2)][-2m\cset{1,2}][-m\cset{1,3}]         $ \\
\vspace{1mm} & $\ca{P}^0 $ & $\to$ & $\{1,1,2\} \to \{2,2\}$ & $\to$ & $\{4\} $ &:&  $ [-6\,c(2)][-c(2)][-m\cset{2,2}]         $ \\
\end{tabular}
\end{equation}
Summing over all maps gives the unitarity relation
\begin{align} \label{eq:s4unitarity}
s\cset{1,1,1,1} &= 1 - 6\,c(2) - 4\,c(3) + 12\,c(2)\,m\cset{1,2} -3\,c(2)^2+ 6\,c(2)^2 - c(4) \nn\\
& \quad + 4\,c(3)\,m\cset{1,3} +3\,c(2)^2\,m\cset{2,2} + 6\,c(2) m\cset{1,1,2} \nn\\
&\quad - 12\,c(2)\,m\cset{1,2}\,m\cset{1,3} - 6\,c(2)^2 m\cset{2,2} \,. 
\end{align}
We give the clustering diagrams for 4 particles in \fig{Case4}, which can alternatively be used to derive this relation.

This map structure is very useful once we add in the measurement on the final state.  There is another important identification to make in these maps.  We can divide a map $\ca{P}^{\de} \to \ca{P}^{f}$ into two parts:
\begin{equation}  \label{eq:phimapproduct}
\begin{tabular}{c c c c c c}
$\phi \, :$  & $ \ca{P}^{\de}$ &  $\quad \xrightarrow[\hspace{1.3cm}]{} \quad$ & $  \ca{P}^{c} $ & $\quad  \xrightarrow[\hspace{2.7cm}]{} \quad$ & $\ca{P}^{f} $ \\ 
& & clustering & & merging/clustering &
\end{tabular}
\end{equation}
as in 
\be \label{eq:PdPcmaps}
\phi(\ca{P}^{\de} , \ca{P}^f) = \phi_{\de}(\ca{P}^{\de},\ca{P}^c) \cdot \phi_c (\ca{P}^c, \ca{P}^f) \,.
\ee
This will provide a key simplification in the measurement function.
The map $\ca{P}^{\de} \to \ca{P}^c$ involves only clustering functions.  The map $\ca{P}^c \to \ca{P}^f$ involves merging or clustering functions, but the first step in the map must be merging so that $\ca{P}^c$ is unambiguously defined.  $\ca{P}^c$ can be equal to $\ca{P}^{\de}$ or $\ca{P}^f$, depending on the map.  To make the role of $\ca{P}^c$ clear, we repeat the 4 particle maps with $\ca{P}^c$ explicitly identified.
\newpage
\newpage
\begin{equation}\label{eq:4MeasClusMapsPc}
\begin{tabular}{c c c c c c c c c c}
\vspace{1mm} $\boldsymbol{\phi:}$ & $\boldsymbol{ \ca{P}^{\de} }$ & & $\boldsymbol {\longrightarrow }$ & & $\boldsymbol{ \ca{P}^c}$ & & $\boldsymbol{\longrightarrow}$ & & $\boldsymbol{\ca{P}^{f}}$ \\
\vspace{1mm} & $\{1,1,1,1\} $ & & $\longrightarrow$ & & $\{1,1,1,1\}$ & & $\longrightarrow$ & & $\{1,1,1,1\}$ \\
\vspace{1mm} & $\{1,1,1,1\} $ & & $\longrightarrow$ & & $\{1,1,2\}$ & & $\longrightarrow$ & & $\{1,1,2\}$ \\
\vspace{1mm} & $\{1,1,1,1\} $ & & $\longrightarrow$ & & $\{1,3\}$ & & $\longrightarrow$ & & $\{1,3\}$ \\
\vspace{1mm} & $\{1,1,1,1\} $ & & $\longrightarrow$ & & $\{1,1,2\}$ & & $\longrightarrow$ & & $\{1,3\}$ \\
\vspace{1mm} & $\{1,1,1,1\} $ & & $\longrightarrow$ & & $\{2,2\}$ & & $\longrightarrow$ & & $\{2,2\}$ \\
\vspace{1mm} & $\{1,1,1,1\} $ & $\to$ & $\{1,1,2\}$ & $\to$ & $\{2,2\}$ & & $\longrightarrow$ & & $\{2,2\}$ \\
\vspace{1mm} & $\{1,1,1,1\} $ & & $\longrightarrow$ & & $\{4\}$ & & $\longrightarrow$ & & $\{4\}$ \\
\vspace{1mm} & $\{1,1,1,1\} $ & & $\longrightarrow$ & & $\{1,3\}$ & & $\longrightarrow$ & & $\{4\}$ \\
\vspace{1mm} & $\{1,1,1,1\} $ & & $\longrightarrow$ & & $\{2,2\}$ & & $\longrightarrow$ & & $\{4\}$ \\
\vspace{1mm} & $\{1,1,1,1\} $ & & $\longrightarrow$ & & $\{1,1,2\}$ &  & $\longrightarrow$ & & $\{4\}$ \\
\vspace{1mm} & $\{1,1,1,1\} $ & & $\longrightarrow$ & & $\{1,1,2\}$ & $\to$ & $\{1,3\}$ & $\to$ & $\{4\}$ \\
\vspace{1mm} & $\{1,1,1,1\} $ & $\to$ & \{1,1,2\} & $\to$ & $\{2,2\}$ & & $\longrightarrow$ & & $\{4\}$ \\
\end{tabular}
\end{equation}
Note that parts of the map where the partition does not change (such as $\{1,1,2\} \to \{1,1,2\}$) contribute nothing to the unitarity relation. 

So far we have only discussed phase space constraints from clustering.  We will now include the effects of the measurement, where much of this structure is employed.  For reference, we give a brief dictionary of symbols defined up to this point in \tab{dictionary}.

\vspace{1em}
\begin{table}[htdp]
\begin{center}
\begin{tabular}{|c|c|}
\hline
\textbf{Symbol} & \textbf{Description} \\
\hline
\hline
$\ca{P}$ & Partition giving the outcome of the algorithm. \\
 & See \eq{Pdef}. \\
\hline
$c(p)$ & Requires $p$ particles to cluster. \\
 & See \eq{cdef}. \\
\hline
$m\cset{p_1,\ldots,p_k}$ & Requires groups of $p_1,\ldots,p_k$ particles to merge. \\
 & See \eq{mdef}. \\
\hline
$s\cset{p_1,\ldots,p_k}$ & Requires groups of $p_1,\ldots,p_k$ particles to not cluster. \\ 
 & See \eq{sdef}. \\
\hline
$\ca{P}^0$ & The base partition $\{1,\ldots,1\}$. \\
 & See \eq{Pmapdefs}. \\
\hline
$\ca{P}^{\de}$ & The partition giving the outcome of the algorithm, used in maps. \\
 & See \eq{Pmapdefs}. \\
\hline
$\ca{P}^{c}$ & A coarser partition than $\ca{P}^{\de}$, obtained by clustering single particles. \\
 & See \eq{4MeasClusMapsPc} for an example. \\
\hline
$\ca{P}^{f}$ & The final step in a map, obtained from $\ca{P}^c$ by merging or clustering. \\
 & See \eq{Pmapdefs}. \\
\hline
$\phi_{\de}(\ca{P}^{\de},\ca{P}^c)$ & A map from $\ca{P}^{\de}$ to $\ca{P}^c$, used in rewriting $s(\ca{P}^{\de})$. \\
 & See \eq{phimapproduct}. \\
\hline
$\phi_c(\ca{P}^{c},\ca{P}^f)$ & A map from $\ca{P}^{c}$ to $\ca{P}^f$, used in rewriting $s(\ca{P}^{\de})$. \\
 & See \eq{phimapproduct}. \\
\hline
\end{tabular}
\end{center}
\caption{Dictionary of symbols defined to describe phase space constraints from clustering.}
\label{tab:dictionary}
\end{table}
%

%%%%%%%%%%%%%%%%%%%%%%%%%%%%%%%%%%%%%%%%%%%%%%%%%%%
\section{The Measurement Function for Clustering Algorithms}
\label{sec:measurement}
%%%%%%%%%%%%%%%%%%%%%%%%%%%%%%%%%%%%%%%%%%%%%%%%%%%

The outcome of a jet algorithm is associated with a measurement on the final state.  Each partition $\ca{P}$ gives a contribution $\cM(\ca{P})$ to the measurement function, where $\cM(\ca{P})$ is the Fourier transform of the measurement function.  To find the total $n$-particle measurement function we can simply weight each term in the unitarity relation of \eq{clusteringunitarity} by the measurement factor for that partition, so that
\be \label{eq:nloopgeneral}
\cM^{(n)} = \sum_{\ca{P}} N(\ca{P}) \bigg[ \prod_{p\in\ca{P}} c(p) \bigg] \cM(\ca{P}) \, s(\ca{P}) \,.
\ee
This is a more specific form of \eq{genMeasK} in transform space.  A similar form of this equation applies to momentum space; with the appropriate mapping $\cM \to \ca{M}$. The measurement function in \eq{nloopgeneral} can be applied to any standard clustering algorithm with any observable. In order to simplify the effect of clustering on the structure of $\cM^{(n)}$ we would like to use the maps developed in \ssec{clusteringmaps} to rewrite $s(\ca{P})$ in terms of clustering and merging functions.  In order to do this we require that the observables that we measure, $\{x\}$ in \eq{genMeasPos}, are only a function of momenta in each subset $p$ and do not depend on multiple subsets in $\ca{P}$.  This is made explicit in \eq{genMeasK}, and hence in \eq{nloopgeneral}, where the phase space constraints from clustering are separated from the measurement\footnote{In principle, one can define observables that can only be written as a function of all the momenta in an event, and cannot be disentangled into contributions from separate regions of phase space.  However, such observables tend not to arise in practical applications.}. This means that the contribution to the measurement of a given outcome $\ca{P}$ of running the jet algorithm can be written as 
\be \label{eq:cdeltaprod}
\cM(\ca{P}) = \prod_{p\in\ca{P}} \cd(p) \,,
\ee
where
\be \label{eq:cMdef}
\cd(p) \text{ : contribution of $p$ clustered particles to the measurement.}
\ee
Note that the single particle measurement function is $\cM^{(1)} = \cd(1)$, so in the absence of clustering 
\be 
\cM\cset{1,\ldots,1} = \Big[\cM^{(1)} \Big]^n \,.
\ee

As an example of a measurement that satisfies \eq{cdeltaprod}, consider running an inclusive jet algorithm and measuring the jet mass for jets with $p_{\rm T}>p_{\rm T}^{\rm cut}$. Each $p \in \ca{P}$ corresponds to a candidate jet produced by running the clustering algorithm, for which we measure its mass $m_J^p$ (or $x_J^p$ in Fourier space). In this case $\cM(\ca{P})$ becomes
\be
\cM(\ca{P}) = \prod_{p\in\ca{P}} \exp \left[ -i \, x_J^p \, \theta\Big(p_{\rm T}(p) >p_{\rm T}^{\rm cut}\Big) \sum_{i,j \in p} q_i \cdot q_j  \right]\,,
\ee
where $q_i$ are the momenta of particles in a given partition $p$. Note that for this observable, the one particle measurement function is $\cM(1) = 1$ for massless final states.  

For the very general class of observables satisfying \eq{cdeltaprod}, the $n$-particle measurement function in \eq{nloopgeneral} becomes 
\be \label{eq:nloopmaster}
\cM^{(n)} = \sum_{\ca{P}} N(\ca{P}) \bigg[ \prod_{p\in\ca{P}} c(p)\, \cd(p) \bigg] s(\ca{P}) \,,
\ee
where all partitions have $n$ particles.  We now show how the maps in \ssec{clusteringmaps} can be used to rewrite $s(\ca{P})$ to express the full measurement function in a form that makes the effect of clustering explicit.

We can think of the measurement as a map from the base partition $\ca{P}^0 = \{1,\ldots,1\}$ to the outcome of the algorithm $\ca{P}^{\de}$:
\be
\text{measurement map } \; \phi_m (\ca{P}^0, \ca{P}^{\de}) \; : \; \ca{P}^0 \; \to \; \ca{P}^{\de} \,.
\ee
Each map is associated with the clustering and measurement factor for the partition.  The full phase space constraints includes a factor of $s(\ca{P})$, but we define the contribution from the map $\phi_m$ without this factor. The $s(\ca{P})$ factors will be rewritten using the maps $\phi_w$ and $\phi_c$ from \ssec{clusteringmaps}.  For 3 particles, the measurement maps and contribution to \eq{nloopmaster} are
\begin{equation}\label{eq:3MeasMaps}
\begin{tabular}{c c c c c c}
\vspace{1mm}
$\boldsymbol{\phi_m\,: }   $&      $\boldsymbol{\ca{P}^0}$ & $\boldsymbol{\to}$ & $\boldsymbol{\ca{P}^{\de} }$ &&  \bf{contribution }    \\
 \vspace{-3mm}\\
\vspace{1mm} & $\{1,1,1\} $ & $\to$ & $\{1,1,1\}    $ &:&  $ \cd(1)^3 $ \\
\vspace{1mm}  & $\{1,1,1\} $ & $\to$ & $\{3\}          $ &:&  $  c(3)\cd(3) $ \\
\vspace{1mm}  & $\{1,1,1\} $ & $\to$ & $\{1,2\}          $ &:&  $  3\,c(2)\cd(2) \cd(1)$ \\
\end{tabular}
\end{equation}
The measurement map corresponds to \eq{nloopmaster} with $\ca{P}$ replaced by $\ca{P}^{\de}$.  The remaining work is to rewrite $s(\ca{P}^{\de})$ in terms of inclusive constraints, which we use the clustering maps for.

Putting the measurement map together with the clustering maps defined in \ssec{clusteringmaps}, we have a complete map from the base partition to the final partition:
\be
\phi_{\rm tot} (\ca{P}^0, \ca{P}^f) = \phi_m (\ca{P}^0 , \ca{P}^{\de}) \cdot \phi_{\de} (\ca{P}^{\de}, \ca{P}^c) \cdot \phi_c (\ca{P}^c, \ca{P}^f) \,.
\ee
If we write all such maps, then we can determine the outcome of the algorithm.

\begin{figure}[t]{
	\begin{center}
	\includegraphics[width=\textwidth]{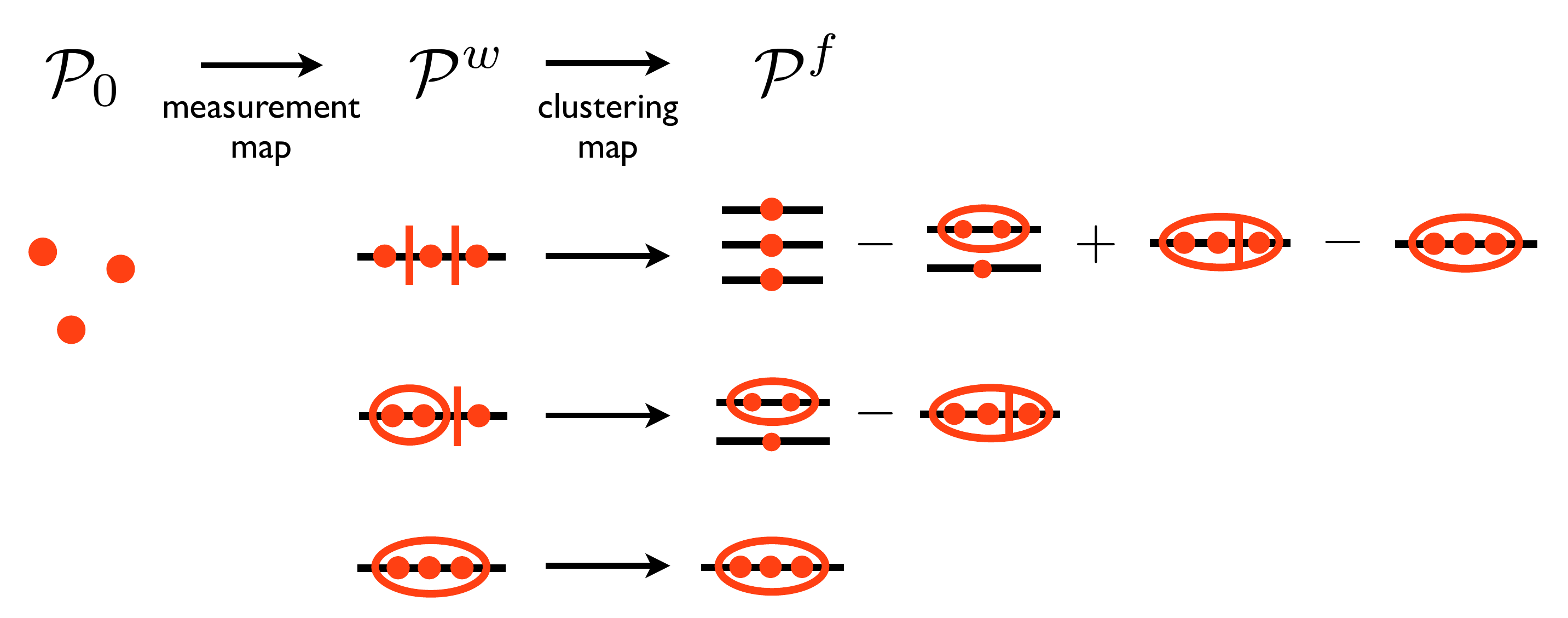} 
	  \end{center} 
	  \vspace{-2em}
{ \caption[1]{The maps to determine the measurement function for 3 particles are shown diagrammatically. The measurement map $\ca{P}^0\to\ca{P}^\de$ corresponds to \eq{3MeasMaps}, while the clustering maps $\ca{P}^\de \to \ca{P}^f$ rewrite $s(\ca{P}^\de)$ using the unitarity relations displayed in \fig{Cases2and3}. The contribution from each map is given in \eq{3MeasClusMaps}. 
   }
  \label{fig:measclusmap}} }
\end{figure}

As an example, we give the measurement functions for the 2 and 3 particle cases. In \fig{measclusmap} we use the diagrammatic notation of \sec{unitarity} to show the measurement maps $\ca{P}^0\to\ca{P}^\de$ and clustering maps $\ca{P}^\de \to \ca{P}^c \to \ca{P}^f$ for the 3 particle case.  
For $n=2$, there are three maps. 
\begin{equation}
\begin{tabular}{c c c c c c c c c c}
\vspace{1mm}
$\boldsymbol{\phi_{\rm tot} : }$      & $\boldsymbol{\ca{P}^0}$ & $\boldsymbol{\to}$ & $\boldsymbol{\ca{P}^{\de}}$ & $\boldsymbol{\to}$ & $\boldsymbol{\ca{P}^{c}}$ & $\boldsymbol{\to}$ & $\boldsymbol{\ca{P}^{f}}$ &\bf{:}&  \bf{contribution}      \\
 \vspace{-3mm}\\
\vspace{1mm} & $\{1,1\} $ & $\to$ & $\{1,1\}        $ & $\to$ & $\{1,1\}   $ & $\to$ & $\{1,1\}   $ &:&  $ \cd(1)^2      $ \\
\vspace{1mm} & $\{1,1\} $ & $\to$ & $\{1,1\}        $ & $\to$ & $\{2\}     $ & $\to$ & $\{2\}     $ &:&  $ -c(2)\cd(1)^2 $ \\
\vspace{1mm} & $\{1,1\} $ & $\to$ & $\{2\}          $ & $\to$ & $\{2\}     $ & $\to$ & $\{2\}     $ &:&  $  c(2)\cd(2)   $ \\
\end{tabular}
\end{equation}
Therefore the total measurement function is
\begin{align} \label{eq:M2}
\cM^{(2)} &= \cd(1)^2 + c(2) \big[ \cd(2) - \cd(1)^2 \big] \nn \\
&\equiv \big[ \cM^{(1)} \big]^2 + \Delta \cM^{(2)} \,,
\end{align}
which defines $\Delta \cM^{(2)}$. For $n=3$, when the measurement maps from \eq{3MeasMaps} are combined with clustering maps, which rewrite $s(\ca{P}^\de)$, there are seven maps.
\begin{equation}\label{eq:3MeasClusMaps}
\begin{tabular}{c c c c c c c c c c}
\vspace{1mm}
$\boldsymbol{\phi_{\rm tot}:}$      & $\boldsymbol{\ca{P}^0 } $ & $\boldsymbol{\to}$ & $\boldsymbol{\ca{P}^{\de}}$ & $\boldsymbol{\to}$ & $\boldsymbol{\ca{P}^{c}}$ & $\boldsymbol{\to}$ & $\boldsymbol{\ca{P}^{f} }       $ &\bf{:}&  \bf{contribution }                     \\
 \vspace{-3mm}\\
\vspace{1mm} & $\ca{P}^0 $ & $\to$ & $\{1,1,1\}      $ & $\to$ & $\{1,1,1\}      $ & $\to$ & $\{1,1,1\}         $ &:&  $ \cd(1)^3                      $ \\
\vspace{1mm} & $\ca{P}^0 $ & $\to$ & $\{1,1,1\}      $ & $\to$ & $\{1,2\}           $ & $\to$ & $\{1,2\}           $ &:&  $ [\cd(1)^3 ][ -3\,c(2)]             $ \\
\vspace{1mm} & $\ca{P}^0 $ & $\to$ & $\{1,2\}        $ & $\to$ & $\{1,2\}           $ & $\to$ & $\{1,2\}           $ &:&  $ 3c(2)\cd(2)\cd(1)             $ \\
\vspace{1mm} & $\ca{P}^0 $ & $\to$ & $\{1,1,1\}      $ & $\to$ & $\{1,2\}$ & $\to$ & $\{3\} $ &:&  $ [\cd(1)^3 ][ -3\,c(2)][-m\cset{1,2}]   $ \\
\vspace{1mm} & $\ca{P}^0 $ & $\to$ & $\{1,2\}        $ & $\to$ & $\{1,2\}        $ & $\to$ & $\{3\}             $ &:&  $ [3c(2) \cd(2)\cd(1)][-m\cset{1,2}] $ \\
\vspace{1mm} & $\ca{P}^0 $ & $\to$ & $\{1,1,1\}      $ & $\to$ & $\{3\}             $ & $\to$ & $\{3\}             $ &:&  $ [\cd(1)^3][-c(3)]                 $ \\
\vspace{1mm} & $\ca{P}^0 $ & $\to$ & $\{3\}          $ & $\to$ & $\{3\}             $ & $\to$ & $\{3\}             $ &:&  $  c(3)\cd(3)                   $ \\
\end{tabular}
\end{equation}
These combined to give the measurement function
\begin{align} \label{eq:M3}
\cM^{(3)} &= \cd(1)^{3} + 3c(2) \big[ \cd(2) - \cd(1)^2 \big] \cd(1) \nn \\
& \qquad \quad  + \Big( c(3)\big[\cd(3) - \cd(1)^3 \big] - 3c(2)m\cset{1,2} \big[\cd(2) - \cd(1)^2 \big] \cd(1) \Big) \nn \\
&\equiv \big[ \cM^{(1)} \big]^3 + 3 \cM^{(1)} \Delta \cM^{(2)} + \Delta \cM^{(3)} \,.
\end{align}
This equation defines $\Delta \cM^{(3)}$.  Note that for the 2 and 3 particle cases, a factor of $\cd(k)$ for $k >1$ comes with an accompanying $-\cd(1)^k$.  This is true in general, which we can show by making use of the separation between the maps $\phi_{\de}$ and $\phi_c$.  This will lead to a simplification of the structure of the measurement function.

Consider the set of maps $\phi_m(\ca{P}^0,\ca{P}^{\de}) \cdot \phi_{\de}(\ca{P}^{\de},\ca{P}^c)$ from $\ca{P}^0$ to $\ca{P}^c$ with a fixed $\ca{P}^c$.  These maps are generated by taking all possible $\ca{P}^{\de}$.  For example, for $n=5$ and $\ca{P}^c = \{1,2,2\}$, these maps are (with $\ca{P}^{\de}$ in bold for each map):
\begin{equation}
\begin{tabular}{ c c c c c c c c}
\vspace{1mm} $\boldsymbol{\phi_m \cdot \phi_{\de} :}$      & $\boldsymbol{\ca{P}^0  }$ & & $\boldsymbol{\longrightarrow}$ & & $\boldsymbol{\ca{P}^{c}}$ &:&  \bf{contribution }                     \\
\vspace{1mm} & ${\bf \{1,1,1,1,1\}}$ & $\to$ & $\{1,1,1,2\}      $ & $\to$ & $\{1,2,2\}      $ &:&  $ 15 \,c(2)^2\cd(1)^5       $ \\
\vspace{1mm} & $\{1,1,1,1,1\}      $ & $\to$ & ${\bf \{1,1,1,2\}}$ & $\to$ & $\{1,2,2\}      $ &:&  $ -30\,c(2)^2\cd(2)\cd(1)^3 $ \\
\vspace{1mm} & $\{1,1,1,1,1\}      $ & $\to$ & $\{1,1,1,2\}      $ & $\to$ & ${\bf \{1,2,2\}}$ &:&  $ 15 \,c(2)^2\cd(2)^2\cd(1) $ \\
\end{tabular}
\end{equation}
The sum of these terms is
\be
15\,c(2)^2 \big[\cd(2) - \cd(1)^2 \big]^2 \cd(1) \,.
\ee
This binomial structure persists.  All terms will share the same clustering factors, differing only in the combinatoric and measurement factors.  If we sum over all $\ca{P}^{\de}$ for a given $\ca{P}^c$ ($\ca{P}^\de \vert \ca{P}^c$), then the clustering and measurement factors are
\be \label{eq:Pcmap}
N(\ca{P}^c) \Big[ \prod_{p\in \ca{P}^c} c(p) \Big] \sum_{\ca{P}^\de \vert \ca{P}^c} N_\de(\ca{P}^{\de}, \ca{P}^c) \Big[ \prod_{p\in \ca{P}^{\de}} \cd(p) \Big] \,.
\ee
The combinatoric factor is defined relative to $N(\ca{P}^c)$, and so it is the number of ways to distinguish the map $\ca{P}^0 \to \ca{P}^{\de} \to \ca{P}^c$ from $\ca{P}^0 \to \ca{P}^c$.  This is simply
\be
N_\de(\ca{P}^{\de}, \ca{P}^c) = \prod_{k=2}^{n} \binom{d_k^c}{d_k^{\de}} \,,
\ee
where $d_k^i$ is the number of elements of $\ca{P}^i$ equal to $k$ (recall \eq{NPdef}).  Therefore, we can sum over all $\ca{P}^{\de}$ and \eq{Pcmap} becomes
\be
N(\ca{P}^c) \Big[ \prod_{p\in \ca{P}^c} c(p) \Delta\cd(p) \Big] \,,
\ee
where
\be \label{eq:Deltadeltadef}
\Delta\cd (p) = \left\{ \begin{split}
&\cd (p) - \cd(1)^{p}  \qquad &\text{if } p > 1 \,, \\
&\cd(1) \qquad &\text{if } p = 1 \,.
\end{split} \right.
\ee

Finally, we must deal with the map $\phi_c(\ca{P}^c, \ca{P}^f)$.  Given a partition $\ca{P}^c$, we will define the set of all maps $\phi_c (\ca{P}^c, \ca{P}^f)$ (all such maps for all allowed $\ca{P}^f$) as $\ca{G}_c (\ca{P}^c)$.  For example,
\be \label{eq:Gcexample}
\ca{G}_c \cset{1,1,2} = \{ \phi_c(\{1,1,2\},\{1,1,2\}), \; \phi_c(\{1,1,2\},\{1,3\}), \; \phi_c(\{1,1,2\},\{4\}) \} \,.
\ee
These maps from $\ca{P}^c$ to $\ca{P}^f$ are the second, third, and eighth maps in \eq{4MeasClusMapsPc}.  For each map $\phi_c$, we define
\be\label{eq:Fdef}
\ca{F}(\phi_c)\, : \, \text{ the product of inclusive constraints for the map } \phi_c  \,.
 \ee
Using the maps in $\ca{G}_c \cset{1,1,2}$ as an example,
\begin{align} \label{eq:Fphiexample}
\ca{F}\big(\phi_c(\{1,1,2\},\{1,1,2\})\big) &= 1 \nn \\
\ca{F}\big(\phi_c(\{1,1,2\},\{1,3\})\big) &= -2\,m\cset{1,2} \nn \\
\ca{F}\big(\phi_c(\{1,1,2\},\{4\})\big) &= 2\,m\cset{1,2}\,m\cset{1,3} \,.
\end{align}

%%%%%%%%%%%%%%%%%%%%%%%%%%%%%%%%%%%%%%%%%%%%%%%%%%%
\subsection{Main Results}
\label{ssec:mainresults}
%%%%%%%%%%%%%%%%%%%%%%%%%%%%%%%%%%%%%%%%%%%%%%%%%%%

Putting the pieces together, the measurement function is
\be \label{eq:MeasFunc}
\boxed{
\ca{M}^{(n)} = \sum_{\ca{P}^c} N(\ca{P}^c) \Big[ \prod_{p\in \ca{P}^c} c(p) \Delta\cd(p) \Big] \sum_{\phi \in \ca{G}_c (\ca{P}^c)} \ca{F} (\phi) \,.
}
\ee
This is one of our main results, and can be used to arrange the measurement function into a form suitable to study the logarithmic contributions to the cross section.

Each map $\phi : \ca{P}^0 \to \ca{P}^f$ will contain some number of single particles not participating in any clustering or merging.  These particles will only be associated with factors of $\cd(1) = \cM^{(1)}$.  Since $\cM^{(1)}$ only depends on the phase space constraints from a single particle, it is useful to distinguish these contributions from multi-particle phase space constraints\footnote{When we consider boundary clustering, in many cases $\cM^{(1)}$ is the only factor that contributes at leading log to the cross section.  For an example, see \cite{Kelley:2012kj}, where the contribution $\Delta \cM^{(2)}$ was calculated for the C/A and $\kt$ algorithms for a jet mass observable in dijet events.  This will be discussed further in \sec{AbelianAllOrders}.}.

Suppose we consider maps with exactly $k$ factors of $\cM^{(1)}$.  These single particles do not participate in the maps at all, and will remain in the final partition $\ca{P}^f$ for each map.  Therefore we can factor out these particles from the map, and they will become maps on $n-k$ particles:
\be
\phi_n \, \to \, \phi_{n-k}
\ee
If the map $\phi_n$ leads to a contribution $\cM (\phi_n)$ to the measurement function, then we can write this as
\be
\cM (\phi_n) = \frac{n!}{(n-k)! \,k!} \, \cM (\phi_{n-k}) \big[ \cM^{(1)} \big]^k \,.
\ee
The combinatoric factor changes the $1/n!$ from the $n$-particle matrix element into separate factors for $\cM (\phi_{n-k})$ and $\big[\cM^{(1)}\big]^k$.  This gives another of our main results,
\be \label{eq:Mnform}
\boxed{
\cM^{(n)} = \big[\cM^{(1)} \big]^n + \sum_{k=0}^{n-2} \frac{n!}{(n-k)! \, k!} \, \Delta \cM^{(n-k)} \big[\cM^{(1)}\big]^k \,,
}
\ee
where $\Delta \cM^{(n)}$ are the $n$-particle measurement function terms where all particles cluster or merge with at least one other particle, and we can use this equation to define $\Delta \cM^{(n)}$. For 2 particles,
\begin{align} \label{eq:DeltaM2}
\Delta \cM^{(2)} &= c(2)\big[ \cd(2) - \cd(1)^2 \big] \nn \\
&= c(2) \Delta \cd(2) \,,
\end{align}
and for 3 particles,
\begin{align} \label{eq:DeltaM3}
\Delta \cM^{(3)} &= c(3)\big[ \cd(3) - \cd(1)^3 \big] - 3\,c(2)m\cset{1,2}\big[ \cd(2) - \cd(1)^2 \big] \cd(1) \nn \\
&= c(3)\Delta \cd(3) - 3\, c(2) m\cset{1,2} \Delta \cd(2) \cd(1) \,.
\end{align}
We give the measurement functions for $n=4$ and $n=5$ in Appendix~\ref{sec:n4Ex} as additional examples.

%%%%%%%%%%%%%%%%%%%%%%%%%%%%%%%%%%%%%%%%%%%%%%%%%%%
 \section{Cancellation of Infrared Divergences from Clustering}
\label{sec:IRdiv}
%%%%%%%%%%%%%%%%%%%%%%%%%%%%%%%%%%%%%%%%%%%%%%%%%%%

The formalism for the measurement function lets us study the IR structure of the various clustering contributions for the Abelian terms.  As an example, we consider the 3 particle terms in the measurement function in the limit that one particle is very soft.  This argument can be generalized to higher order terms, or adapted to the collinear limit.  For 3 particles,
\begin{align} \label{eq:M3terms}
\cM^{(3)} &= \big[ \cM^{(1)} \big]^3 + 3 \Delta\cM^{(2)} \cM^{(1)} + \Delta\cM^{(3)} \nn \\
&= \cd(1)^3 + 3 c(2) \big[\cd(2) - \cd(1)^2\big] \cd(1) \nn \\
& \qquad + c(3) \big[\cd(3) - \cd(1)^3\big] - 3c(2)m\cset{1,2}\big[\cd(2) - \cd(1)^2\big] \cd(1) \,.
\end{align}
The single particle measurement function $\cM^{(1)}$ leads to an IR safe contribution to the cross section.  We will consider $\Delta \cM^{(3)}$, which is the last line of \eq{M3terms}; in showing these terms are IR safe, we will show that terms involving $\Delta \cM^{(2)}$ are also free of IR divergences.

Let us label the soft momenta $k_1$, $k_2$, and $k_3$; we will take the limit $k_3 \to 0$.  The term involving the constraints $c(2) m\cset{1,2}$ has been symmetrized over soft momenta, meaning if we want to take $k_3$ soft then we must split this up into 3 separate terms.  Labeling the momenta, this term becomes
\begin{align} \label{eq:m12terms}
&3c(2)m\cset{1,2} \big[\cd(2) - \cd(1)^2 \big] \cd(1) \nn \\
& \qquad \to \, c\cset{k_1,k_2} m\cset{k_3,\{k_1,k_2\}} \big[ \cd(k_1,k_2) - \cd(k_1)\cd(k_2) \big] \cd(k_3) \nn \\
& \qquad \qquad + c\cset{k_1,k_3} m\cset{k_2,\{k_1,k_3\}} \big[ \cd(k_1,k_3) - \cd(k_1)\cd(k_3) \big] \cd(k_2) \nn \\
& \qquad \qquad + c\cset{k_2,k_3} m\cset{k_1,\{k_2,k_3\}} \big[ \cd(k_2,k_3) - \cd(k_2)\cd(k_3) \big] \cd(k_1) \,.
\end{align}
In this language, the IR safety of $\Delta \cM^{(2)}$ is due to the relations
\be \label{eq:cdeltaIR}
\lim_{k_s \to 0} \cd(k,k_s) = \cd(k) \,, \qquad \lim_{k_s \to 0} \cd(k_s) = 1 \,.
\ee
This implies that 2 particle clustering effects are IR safe:
\be
\lim_{k_s \to 0} \cd(k,k_s) - \cd(k)\cd(k_s) = 0 \quad \Rightarrow \quad \lim_{k_s \to 0} \Delta \cM^{(2)} = 0 \,.
\ee
This relation simplifies the terms in \eq{m12terms} to
\be
c\cset{k_1,k_2} m\cset{k_3,\{k_1,k_2\}} \big[ \cd(k_1,k_2) - \cd(k_1)\cd(k_2) \big] \,.
\ee
The relation in \eq{cdeltaIR} can be generalized to
\be
\lim_{k_s \to 0} \cd(n) \to \cd(n-1) \,,
\ee
which means that the $c(3)$ term in $\Delta \cM^{(3)}$ becomes
\be \label{eq:c3term}
c\cset{k_1,k_2,k_3} \big[\cd\cset{k_1,k_2} - \cd(k_1)\cd(k_2) \big] \,.
\ee
Therefore both groups of terms are IR divergent in the soft limit.  However, by considering the merging constraints one can show for an IR safe jet algorithm that
\be
\lim_{k_3 \to 0} c\cset{k_1,k_2} m\cset{k_3,\{k_1,k_2\}} = c\cset{k_1,k_2,k_3} \,.
\ee
The reason is that when $k_3$ is soft, clustering with any other particle $k_i$ would leave the momentum of $k_i$ unchanged.  In this limit the constraints requiring the soft particle to merge with a clustered pair is the same as all three particles clustering.  Therefore, both terms in $\Delta \cM^{(3)}$ reduce to \eq{c3term}, and they cancel ($\Delta\cM^{(3)} \to 0$).  This is an explicit demonstration that IR divergences are not introduced by clustering effects for 2 or 3 particles.

%%%%%%%%%%%%%%%%%%%%%%%%%%%%%%%%%%%%%%%%%%%%%%%%%%%
\section{An Application: All-Orders Abelian Structure of Soft Clustering for Jet Shapes}
 \label{sec:AbelianAllOrders}
%%%%%%%%%%%%%%%%%%%%%%%%%%%%%%%%%%%%%%%%%%%%%%%%%%%

The all-orders description of the measurement function developed in the previous sections is useful because it enables us to understand the effects of clustering on the higher order terms in the cross section for jet observables. As an example we consider $e^+e^-\to N$ jets defined by an inclusive jet algorithm (such as C/A) and define two observables, $\rho$ and $\Lambda$, where $\rho = (m_1^2 + \ldots + m_N^2) / Q^2$ is the sum of jet masses $\{m_i\}$ scaled by the center of mass energy $Q$ and $\Lambda$ is the total energy of radiation outside of all jets. When $\{Q \rho,\Lambda\} \ll Q \sqrt{\rho} \ll Q$, each jet mass $m_i$ is forced to be small and the event is described by collinear radiation in the jets and soft radiation in and between the jets. This observable was considered for dijet events in \cite{Banfi:2010pa,KhelifaKerfa:2011zu,Kelley:2012kj,Kelley:2011aa,Kelley:2011tj,Hornig:2011tg}. The all-orders structure of the Abelian terms in this section apply to a wide range of jet shapes in the soft/collinear regime, such as angularities.

Perturbative calculations generate logarithms of the form $\ln \rho$ and $\ln \Lambda/Q$, and a reliable prediction in this regime requires resummation.  This is achieved through a factorization theorem, for which for the double cumulant takes the form
\be \label{eq:cumulantdef}
\Sigma(\rho_c,\Lambda_c) = \int_0^{\rho_c} d\rho \int_0^{\Lambda_c} d\Lambda \, \frac{d^2 \sigma}{d\rho\, d\Lambda} \,,
\ee
where
\be
\label{eq:fact}
\frac{d^2\sigma}{d\rho\, d\Lambda} = \sigma_0 \, H_N(Q) \, \Big[ J_1 (\rho) \otimes \cdots \otimes J_N (\rho) \Big] \otimes S_N (\rho, \Lambda) \,.
\ee
The hard function $H$ is determined by the short distance interaction that produces the high energy jets, and depends on the center of mass energy $Q$.  The jet functions parameterize collinear radiation for each jet at the scale $Q\sqrt{\rho}$ and the soft function parameterizes soft radiation in and between jets at the scale $Q\rho$.  The radiation collinear to each jet direction clusters only within the jet when $\sqrt{\rho} \ll R$, and in this case the effect of clustering is described entirely by the \textit{boundary clustering} of soft particles in the event, as discussed in \sec{intro}.  It was shown in \cite{KhelifaKerfa:2011zu,Kelley:2012kj} that boundary clustering gives rise to clustering NGLs in the soft function of the form $\ln (Q\rho/\Lambda)$ starting at $\ca{O}(\as^2 \ln^2)$.

\be\label{eq:Sfun}
S_N( \rho,\Lambda) = \frac{1}{c_N}  \,\langle 0 \lvert \widehat{\ca{W}}^{\,\dagger}_{\{ n\}} \, \widehat{\ca{M}}( \{ \rho,\Lambda \} )\, \widehat{\ca{W}}_{\{ n\}} \, \rvert 0 \,\rangle 
\ee
where $\widehat{\ca{W}}_{\{ n \}}$ represents a product of soft Wilson lines, $Y_{n_i}$, where $n_i^\mu = (1, \hat{n}_i)$ is a light like vector in the $i$-th jet direction, with the appropriate color representation and path ordering of the final state hard partons\footnote{Recall a soft Wilson line directed along $n^\mu$, in representation $\textbf{T}^a$, contains soft gauge fields ($A_s^\mu$) and is defined by a path ordered exponential:
\begin{equation*}
Y_n^{\dagger}(x) = {\rm P} \exp \left( ig \int_0^{\infty} \!\! ds\  n \cdot A^a_{\rm s}( x + s n) \textbf{ T}^a \right) \,. 
\end{equation*}
}. The soft function is a matrix in color space and $c_N$ is the color normalization factor such that $S_N(\rho,\Lambda)= \delta(\rho)\delta(\Lambda)$ at tree-level. The measurement operator $\widehat{\ca{M}}( \rho,\Lambda )$ implements the measurement of $\rho$ and $\Lambda$ for a given jet algorithm, and, for final states $\ket{X_s}$ with soft momenta $\{ k \}$, the measurement function acts as
\be
\widehat{\ca{M}}( \rho,\Lambda ) \, | X_s \rangle = \ca{M} \big( \rho,\Lambda , \{k \} \big) \, | X_s \rangle \,.
\ee
For convenience we work with the Fourier space soft function,
\be
\wt{S}_N(x,y) =  \int_{-\infty}^{\infty} d\rho \, d\Lambda \, \exp( -ix\rho) \exp(- iy\Lambda/Q) S_N( \rho,\Lambda)\,,
\ee
for which the measurement function is typically expressed in the form
\be \label{eq:MeasPos}
\cM \big( x , y  , \{k\} \big) = \prod_{i=1}^{N} \exp\left( -i x \sum_{j} \frac{n_i\cdot k_j}{Q} \R_{J_i} \big(k_j; \{k\} \big) \right)  \exp\left( -i y\sum_{j} \frac{k_j^0}{Q} \R_{\rm out} \big(k_j; \{k\} \big) \right) \,,
\ee
where the functions $\R_{J_i}\big(k_j; \{k\} \big)$ and $\R_{\rm out}\big(k_j; \{k\} \big)$ implement the phase space constraints of the jet algorithm, requiring particle $k_j$ to be in the $i$-th jet and out-of-jet regions respectively.  Each region, $\R_i$, depends on all the soft momenta in the final state $\{k\}$ and therefore implicitly on all of the other regions.  The formalism developed in this paper helps disentangle this dependence.  For $n$ final state particles it is useful to organize the effects of clustering as in \eq{Mcorr}. 
When the measurement function is integrated against the squared matrix element from the Wilson lines, we get the $\ca{O}(\as^n)$ contribution to the soft function,
\be \label{eq:Snform}
\wt{S}^{(n)} ( \{ x \} )  = \int \left(\prod_{i=1}^n \frac{d^4 k_i}{(2\pi)^4} \right) \ca{A}^{(n)} \big(\{k\}\big) \cM^{(n)} \big(\{ x \} , \{k\} \big) \,.
\ee
The first term in \eq{Mcorr} is the measurement function for an algorithm for which the measurement function factorizes into single particle measurements in the soft sector, such as for the anti-$\kt$ algorithm.  The second term accounts for correlations in the measurement function.  This leads us to write
\be \label{eq:Sresult}
\wt{S}^{(n)} ( \{ x \}  ) = \wt{S}^{(n)}_{\akt} (\{ x \}  ) +   \wt{S}^{(n)}_{\rm corr.} (\{ x \}  ) \,.
\ee
The algorithm-dependent effects of soft clustering are entirely contained in the last term.

When soft gluons boundary cluster they can not be collinear to the jet direction, as depicted in \fig{BoundaryClusEx}. In general a final state gluon can contribute a double log to the cumulant cross section, $\as \ln^2 (Q \rho_c/\Lambda_c)$, associated with soft and collinear divergences. For the Abelian terms in the soft matrix element, collinear divergences arise only for soft gluons collinear to the jet direction. Therefore, as shown in \cite{Kelley:2012kj}, those gluons involved in boundary clustering can each contribute at most a single log, $\as  \ln (Q \rho_c/\Lambda_c)$, from a soft divergence to the Abelian terms in $\wt{S}_{\rm corr}^{(n)}$. For example, at $\ord{\as^3}$, $\cM^{(3)}_{\rm corr}$ contains a contribution where all three particles boundary cluster together, which contribute at NLL to $\wt{S}_{\rm corr}^{(3)}$. However the measurement function also contains the contribution
\begin{align}\label{eq:M3prob}
\cM^{(3)} \supset & \,  \theta \left( \text{gluon 1 and 2 boundary cluster}\right)  \nn\\
& \times \theta\left( \text{gluon 3 is in the jet and does not boundary cluster}\right)\,,
\end{align}
which a priori could contribute at LL in the exponent, $\as^n \ln^m$ for $m >n$, since gluon 3 is sensitive to both soft and collinear divergences. In order to show that clustering NGLs do in fact occur at NLL, we must show that constraints of the form \eq{M3prob} can be expressed in terms of lower order phase space constraints.  Rewriting the measurement function in terms of inclusive constraints allows us to do precisely this. For the term in \eq{M3prob} this is illustrated graphically is \fig{q3R12}.  Next we will use our main result, \eq{Mnform}, to show the all-orders form of the Abelian soft function. We will see that clustering NGLs give new contributions at NLL at all orders, spoiling the exponentiation of the Abelian terms.

%-----------------------------------------------------------------------------------------------------------
\subsection{All-Orders Abelian Structure}
%-----------------------------------------------------------------------------------------------------------
A considerable simplification happens in the Abelian sector.  It is well known that the $\ca{O}(\as^n)$ Abelian matrix element factorizes as
\be \label{eq:Afact}
\ca{A}^{(n)} \big(\{k_s\}\big) = \frac{1}{n!}\prod_{i=1}^n \ca{A}^{(1)} (k_i) \,,
\ee
where $\ca{A}^{(1)} (k)$ is the squared matrix element at $\ca{O}(\as)$.  When clustering effects are absent or power suppressed, the measurement function factorizes into the product of single gluon constraints, as for the anti-$\kt$ algorithm in \eq{Mfact}. In this case it is easy to see that Abelian soft function exponentiates. The  $\ca{O}(\as^n C_F^n)$ contribution to the soft function is\footnote{In pure dimensional regularization, Abelian matrix elements with a virtual gluon are scaleless, hence we only consider the matrix element where every particle is in the final state.}
\begin{align} \label{eq:Sfact}
\wt{S}^{(n)}_{\text{Abel.}} ( \{ x \}  )&= \frac{1}{n!} \left(\int \frac{d^4 k}{(2\pi)^4} \, \ca{A}^{(1)} (k) \, \cM^{(1)} ( \{ x \}  )\right)^n \nn \\
&= \frac{1}{n!} \left[ \wt{S}^{(1)} ( \{ x \}  ) \right]^n \,,
\end{align}
which is summed to produces the-all orders Abelian soft function, 
\be
\label{eq:Sexp}
\wt{S}_{\text{Abel.}} ( \{ x \}  ) = \exp\Big[ \wt{S}^{(1)} ( \{ x \}  ) \Big] \,.
\ee
This is a manifestation of Abelian exponentiation and we see that factorization of the measurement function, as in \eq{Mfact}, is  a necessary condition for it to hold.

For algorithms for which the measurement function contains corrections from clustering, we can use our main result, \eq{Mnform}, to express these corrections as
\be
\cM_{\rm corr.}^{(n)} \big( \{ x\}, \Phi \big) = \sum_{k=0}^{n-2} \frac{n!}{(n-k)! \, k!} \, \Delta \cM^{(n-k)} \big[\cM^{(1)}\big]^k \,,
\ee
where $\Delta \cM^{(n)}$ is the part of the $n$-particle measurement function in which all $n$ particles cluster with at least one other particle.  These corrections modify \eq{Sfact} to give
\be \label{eq:SnformMod}
\wt{S}^{(n)}_{\text{Abel.}}  = 
\frac{1}{ n!}\bigl( \wt{S}^{(1)} \bigr)^n
+
\sum_{k=0}^{n-2} \Delta \wt{S}_{\alg}^{(n-k)} 
\frac{1}{ k!} \bigl( \wt{S}^{(1)} \bigr)^k  \,,
\ee
where we have defined
\begin{align} \label{eq:SfactMod}
\Delta \wt{S}^{(n)}_{\text{alg}} ( \{ x \}  )&=  \frac{1}{n!} \prod_{i=1}^{n} \int \frac{d^4 k_i}{(2\pi)^4} \, \ca{A}^{(1)} (k_i) \, \Delta \cM^{(n)} ( \{ x \} , \{ k \})  \,.
\end{align}
Finally, although the Abelian soft function no longer exponentiates, the form of \eq{SnformMod} allows us to write the all orders soft function as
\be
\label{eq:SNexp}
\wt{S}_{\text{Abel.}} ( \{ x \}  ) = \exp\Big[ \wt{S}^{(1)} ( \{ x \}  ) \Big] \left( 1 + \sum_{k=2}^{\infty}\Delta \wt{S}^{(k)}_{\alg} ( \{ x \}  )  \right) \, .
\ee
Since $\Delta \wt{S}^{(k)}_{\alg}$ arises from phase space constraints where each of the $k$ gluons boundary clusters with \textit{at least} one other gluon, it contributes at most at NLL; that is $\as^k \ln^k (Q\rho_c/\Lambda)$ in the cumulant cross section. Resummation of clustering NGLs would require a relation between $\Delta \wt{S}^{(n)}_{\alg}$ for all $n$. However, as we have shown from the general form the measurement function in \eq{MeasFunc} and \eq{Mnform}, at each order in $\as^n$ the clustering terms in the soft function receive a contribution from 
\be
\Delta \cM^{(n)} \supset c(n) [\cd(n)-\cd(1)^n] \,,
\ee
where all $n$ particles cluster with each other. This contribution is new and unique at this order in $\alpha_s$. This makes resummation of clustering NGLs very unlikely. In the next subsection, we will see that the all-orders form of the Abelian soft function in \eq{SNexp} is required in order to have a renormalizable soft function.

%--------------------------------------------------------------------------------------------
\subsection{Renormalization of the Abelian Soft Function}
%--------------------------------------------------------------------------------------------
It is well known that the renormalization of the product of Wilson lines, such as those that appear in the soft function, obey Abelian exponentiation -  that is, the Abelian contribution to the all-orders anomalous dimension depends only on the the one-loop result.  However,  the presence of the measurement operator in the soft function makes its structure and renormalization more complex.  We now consider the effects that clustering has on the all orders UV structure of the Abelian soft function.

We have seen in \eq{Sexp} that since the anti-$\kt$ measurement function can be expressed as the product of single gluon constraints, the soft function in this case obeys Abelian exponentiation. This means that the bare anti-$\kt$ soft function is related to the renormalized one to all orders as follows:
\begin{align}\label{eq:SaktBare}
\wt{S}^{\,b}_{\akt} = \wt{Z}_{\akt} \, \wt{S}^{\,r}_{\akt}  = \exp\left[\wt{Z}^{(1)}_{\akt} + \wt{S}^{\,r\,(1)} \right] \,,
\end{align}
where $b$ stands for bare, $r$ stands for renormalized, and $\wt{Z}^{(1)}_{\akt}$ is the Abelian position space one-loop counter-term.  We have suppressed the subscript `Abel.'  since all of our statements in this section only refer to Abelian terms. Consistency of the factorization in \eq{fact} and the requirement that the cross section must be renormalization scale independent imply that the anomalous dimensions must satisfy
\be\label{eq:consistency}
\gamma_H + \sum_{i=1}^{N}\gamma_{J_i} + \gamma_S= 0 \,.
\ee
Since the hard and jet function anomalous dimensions do not depend on the algorithm, the soft function anomalous dimension must also be independent of it.  This implies that $\wt{Z}^{(1)}_{\rm alg.} = \wt{Z}^{(1)}_{\akt}$ must hold for all algorithms.  Any contribution to the soft function $\Delta \wt{S}_\alg$ must be UV finite, since no new divergences are introduced by these algorithms when compared to anti-$\kt$.  

Since algorithms in the $\kt$ class share the same $\wt{Z}$ factor,  the bare soft function is related to the renormalized one as follows:
\begin{align}\label{eq:SalgBare}
\wt{S}^{\,b}_{\alg} & = \wt{Z}_{\akt} \, \wt{S}^{\,r}_{\alg} = \exp\left[ \wt{Z}^{(1)}_{\akt} \right]   \, \wt{S}^{\,r}_{\alg} \,,
\end{align}
using \eq{SaktBare}.  When $\wt{S}^{(1)}$ is replaced with its bare version in \eq{SnformMod}, it takes the form
\be
\wt{S}^{\,b\,(n)}_\alg = 
\frac{1}{n!} \bigl(\wt{S}^{\,b\,(1)} \bigr)^n 
+ 
\sum_{k=0}^{n-2}  \Delta \wt{S}^{(n-k)} 
\frac{\bigl( \wt{S}^{\,b\,(1)} \bigr)^k}{ k!}   \,.
\ee
We can rewrite this to all orders as
\begin{align}
\wt{S}^{\,b}_\alg &= \exp \left[\wt{S}^{\,b\,(1)} \right] \left(1+\sum_{k=2}^{\infty} \Delta \wt{S}^{(k)}_{\rm alg.} \right) \,.
\end{align}
After inserting $\wt{S}^{\,b\,(1)} = \wt{Z}^{(1)}_{\akt}  +\wt{S}^{\,r\,(1)} $ into the previous equation, it follows from \eq{SalgBare}, that the renormalized soft function is 
\be \label{eq:SfuncAllOrders}
\wt{S}^{\,r}_\alg = \exp\left[\wt{S}^{\,r\,(1)} \right] \left(1+\sum_{k=2}^{\infty} \Delta \wt{S}^{(k)}_{\rm alg.} \right) \,.
\ee

Lastly, we also know from \sec{IRdiv} that $\Delta \wt{S}^{(n)}_{\rm alg.}$ is IR finite in addition to UV finite.  As such, these terms can be calculated numerically without the need for difficult subtraction schemes.   This is a major advantage, since the phase space constraints imposed by $\Delta \cM^{(n)}$ are typically quite complex, making the effects difficult to calculate analytically.  The $\ca{O}(\as^2)$ corrections for the C/A and $\kt$ algorithms were calculated in \cite{Kelley:2012kj} for the dijet case ($N=2$).

%--------------------------------------------------------------------------------------------
\subsection{Failure of Exponentiation}
%--------------------------------------------------------------------------------------------
We might hope that the corrections due to clustering in the measurement function could be redefined to reveal further structure in the  measurement function beyond that in \eq{Mnform}.  In particular, one might expect that the corrections to the measurement function could be redefined in such a way that the Abelian corrections exponentiate:
\be
\wt{S}^{\,r}_\alg = \exp\left[\wt{S}^{\,r\,(1)} +\sum_{n=2}^{\infty} \Delta \wt{S}\,'^{\,(n)} \right] \,.
\label{eq:DefPrime}
\ee 
This is equivalent to reorganizing the measurement function to define a new correction term, $\Delta \cM\,'^{\,(n)}$ that is integrated against the Abelian matrix element to get $ \Delta \wt{S}\,'^{\,(n)}$, as in \eq{SfactMod}.  The form in \eq{DefPrime} follows from the properties of clustering in the measurement function if we can define $\Delta \wt{S}\,'^{\,(n)}$ at each order to involve only the clustering/merging of \text{all} $n$ particles with each other. This is contrast with the form in \eq{SfuncAllOrders}, where $\Delta \wt{S}^{(n)}$ involves clustering effects of $n$ \emph{or fewer} particles. Of course, we could simply define $\Delta \wt{S}'$ through \eq{DefPrime} at each order, since the correction terms $\Delta \wt{S}^{(k)}_{\rm alg}$ in \eq{SfuncAllOrders} are finite. However, unless we show \eq{DefPrime} from the properties of clustering in the measurement function, it has no more content than \eq{SfuncAllOrders}. 

Terms grouping all $n$ particles together arise, in the language of \sec{measurement}, from mappings that have $\ca{P}^f = \{n\}$.  To illustrate the difference with an explicit example, consider $n=4$, which is worked out in full detail in Appendix~\ref{sec:n4Ex}.  $\Delta \wt{S}^{(4)}$ contains clustering effects of 4 particles as well as clustering effects from 2 particles in two separate groups:
\be \label{eq:DeltaM4prime}
\Delta \cM^{(4)} = 3\big[\Delta \cM^{(2)}]^2 + \Delta \cM\,'^{\,(4)} \,.
\ee
In Appendix~\ref{sec:n4Ex}, we see that $\Delta \cM^{(4)}$ can be rewritten such that every term in 
$\Delta \cM\,'^{\,(4)}$ merges or clusters all 4 particles into one group and is given in \eq{DeltaM4}.  
In general, this approach requires defining  $\Delta \cM\,'^{\,(n)}$ such that clustering effects at lower orders ($\Delta \cM\,'^{\,(k)}$ for $k < n$) are explicitly factored out. However, for $n=5$ (and higher), this is not the case, as demonstrated in Appendix~\ref{sec:n4Ex}.  With 5 particles, there is a leftover term that cannot be placed into products of $\Delta \cM\,'^{\,(k)}$ for $k<5$, nor does it cluster or merge all $5$ particles together.  This term must reside in $\Delta \cM\,'^{\,(5)}$:
\be
\Delta \cM\,'^{\,(5)} \supset 30 \, c(2)^2 m\cset{1,2} \big[\cd(2) - \cd(1)^2\big] \cd(1)^3 \,.
\ee
It contains two groups, one of 2 particles and one of 3 particles. Such terms are associated with the map from $\ca{P}^c = \{1,1,1,2\} \to \ca{P}^f = \{2,3\}$,  when clustering takes place after the first merging step.  For $n \geq 5$, there are terms contributing to $\Delta \cM'^{(n)}$ in which each of the $n$ particles cluster or merge with at least one other particle but all $n$ are not grouped together, i.e. $\ca{P}^f \neq \{ n\}$.   We are forced to conclude that \eq{DefPrime} does not have any content and is simply a redefinition of \eq{SfuncAllOrders}.

%%%%%%%%%%%%%%%%%%%%%%%%%%%%%%%%%%%%%%%%%%%%%%%%%%%
\section{Conclusions}
\label{sec:conclusions}
%%%%%%%%%%%%%%%%%%%%%%%%%%%%%%%%%%%%%%%%%%%%%%%%%%%

In this paper, we developed a formalism to determine the phase space constraints from a clustering algorithm and the effects on a measurement function.  This framework can be used to study perturbative corrections that arise from clustering. We expressed the constraints from clustering as a function of the outcome of the algorithm. This is characterized in terms of a partition that specifies which particles are clustered by a clustering function, $c(p)$, and a non-clustering function, $s(\{p\})$, which requires groups of particles not to cluster. We derived a set of unitarity relations on phase space, which can be iteratively applied to eliminate the non-clustering function.  As discussed in the text, this requires the introduction of a merging function, $m(\{p\})$, which appears in unitarity relations involving $s$ functions.  This rewriting eliminates exclusive phase space constraints that require particles \textit{not} to be in certain regions of phase space, producing only inclusive constraints. This is useful because it simplifies the structure of the most divergent regions of phase space and is a key step in allowing us to relate higher order clustering effects to lower order constraints.

In \sec{measurement}, we applied this result to the measurement function, weighting the phase space constraints for each outcome of the algorithm by the contribution to the measurement.  Grouping terms that have the same phase space constraints from clustering, we rewrote the measurement function purely in terms of inclusive constraints in \eq{MeasFunc}, eliminating the non-clustering function $s$.  This form makes the clustering effects explicit, expressing the measurement function as a term associated with no clustering and a set of corrections.  These correction terms are associated with $n$-particle clustering effects, and are proportional to factors defined in \eq{Deltadeltadef} that are the difference between the contribution to $\cM$ with and without clustering.  This made the analysis in \sec{IRdiv} of the infrared structure associated with clustering more straightforward.  Additionally, we showed that the $n$ particle measurement function $\cM^{(n)}$ can be related to the measurement function appearing at lower orders, $\cM^{(k)}$ with $k<n$, as in \eq{Mnform}. This form is particularly useful when trying to understand the higher order structure of jet cross sections.

In \sec{AbelianAllOrders} we apply our formalism for the measurement function to examine the perturbative corrections that arise from clustering for a wide class of jet shapes in the soft collinear regime. The main results for the measurement function, chiefly \eq{MeasFunc} and \eq{Mnform}, were used to determine the all-orders structure of the Abelian terms in the soft function, \eq{SfuncAllOrders}. This structure is universal to jet shape observables and depends only on the form of the measurement function and constraints from renormalization.  From the all-orders form of the soft function and by count logarithms arising from clustering in the Abelian terms, we found that to all orders clustering NGLs contribute at NLL plus subleading logs for the $\kt$ class of algorithms.  This same framework can be applied to study non-Abelian terms or non-logarithmic contributions from clustering, since the measurement function contributions from clustering are made explicit.

%%%%%%%%%%%%%%%%%%%%%%%%%%%%%%%%%%%%%%%%%%%%%%%%%%%
\section{Acknowledgments}
%%%%%%%%%%%%%%%%%%%%%%%%%%%%%%%%%%%%%%%%%%%%%%%%%%%

We would like to thank Christian Bauer, Christopher Lee, and Matthew Schwartz for helpful comments on a draft of this work.  This work was supported in part by the Office of High Energy Physics of the U.S.\ Department of Energy under the Contract DE-AC02-05CH11231 and Grant DE-SC003916. JW was supported in part by a LHC Theory Initiative Postdoctoral Fellowship, under the National Science Foundation grant PHY-0705682.

%%%%%%%%%%%%%%%%%%%%%%%%%%%%%%%%%%%%%%%%%%%%%%%%%%%
\appendix
%%%%%%%%%%%%%%%%%%%%%%%%%%%%%%%%%%%%%%%%%%%%%%%%%%%

%%%%%%%%%%%%%%%%%%%%%%%%%%%%%%%%%%%%%%%%%%%%%%%%%%%
\section{Examples for 4 and 5 Final State Particles}
\label{sec:n4Ex}
%%%%%%%%%%%%%%%%%%%%%%%%%%%%%%%%%%%%%%%%%%%%%%%%%%%
In this appendix we present the measurement function for $4$ and $5$ final state particles using the formalism of \sec{unitarity} and \sec{measurement}. 
We will start with $4$ final state particles, for which there are five possible partitions.
\begin{equation}
\{ \{4\}, \{1,3\},\{2,2\},\{1,1,2\}, \{1,1,1,1\} \} \,.
\end{equation}
As dictated in \eq{MeasFunc}, we split the contribution into two factors:  the one associated with $\ca{P}^{c}$ and the one associated with the merging sequence. 
For each allowed $\ca{P}^c$, the appropriate factors from the first half of \eq{MeasFunc} are: 
%-------------------------------------------------------------------------------------
\begin{equation}
\begin{tabular}{ c c c  } 
             $\boldsymbol{\ca{P}^c} $ & \bf{: }& $\boldsymbol{ N(\ca{P}^c)  \prod_{p^c \in \ca{P}^c} c(p^c) \Delta \cd (p^c) } \vspace{1mm}  $ \\ \vspace{-3mm} \\
\vspace{2mm} $\{4\}$     & : & $c(4)   \Delta\cd(4)                  $  \\
\vspace{2mm} $\{1,3\}$   & : & $4c(3)   \cd(1)\Delta \cd(3)          $  \\
\vspace{2mm} $\{2,2\}$   & : & $3c(2)^2 \Delta\cd(2)^2               $  \\
\vspace{2mm} $\{1,1,2\}$ & : & $6c(2)   \cd(1)^2\Delta \cd(2)  $  \\
\vspace{2mm} $\{1,1,1,1\}$ & : & $ \cd(1)^4 \,. $  \\
\end{tabular}
\label{eq:n4c}
\end{equation}
%-------------------------------------------------------------------------------------
Each of these factors is then multiplied by $ \sum_{\phi \in \ca{G}_n (\ca{P}^c)} \ca{F} (\phi) $.
The function $\ca{F}(\phi)$, defined in \eq{Fdef}, depends on the specific sequence of mergings that is represented by the mapping
\begin{equation}
\phi_c: \ca{P}^c =  \ca{P}^{\phi}_{1} \to \ca{P}^{\phi}_{2}  \cdots  \ca{P}^{\phi}_{j} \to \ca{P}_{j+1}^{\phi} = \ca{P}^{f}\,.
\end{equation}
Recall, from the definition of this map, that the first step in this mapping, $\ca{P}^{\phi}_{1} \to \ca{P}^{\phi}_{2}$, must be a merging step and not a clustering clustering.  All of the possible mappings for $n = 4$ and the associated factors are listed below.  
%-------------------------------------------------------------------------------------
\begin{equation}
\begin{tabular}{c c c c c } 
 $\boldsymbol{ \phi_c: \ca{P}^c    }    $ & $ \xrightarrow[\hspace{1.0cm}]{}                         $ & $ \boldsymbol{ \ca{P}^f }$ &\bf{:}&  $\boldsymbol{ \ca{F} (\phi) }$ \vspace{1mm} \\    \vspace{-3mm}\\
\vspace{3mm} $\{4\}      $ & $ \xrightarrow[\hspace{1.0cm}]{}                          $ & $ \{4\}     $ &:&  $1                         $ \\
\vspace{2mm} $\{1,3\}    $ & $ \xrightarrow[\hspace{1.0cm}]{}                          $ & $ \{4\}     $ &:&  $- m(\{1,3\})              $ \\
\vspace{2mm} $\{2,2\}    $ & $ \xrightarrow[\hspace{1.0cm}]{}                          $ & $ \{4\}     $ &:&  $- m(\{2,2\})              $ \\
\vspace{2mm} $\{1,1,2\}  $ & $ \xrightarrow[\hspace{1.0cm}]{}                          $ & $ \{4\}     $ &:&  $- m(\{1,1,2\})            $ \\
\vspace{2mm} $\{1,1,2\}  $ & $ \longrightarrow \quad  \{1,3\} \quad \longrightarrow  $ & $ \{4\}     $ &:&  $ 2m(\{1,2\})m(\{1,3\})    $ \\
\vspace{2mm} $\{1,3\}    $ & $ \xrightarrow[\hspace{1.0cm}]{}                          $ & $ \{1,3\}   $ &:&  $ 1                        $ \\
\vspace{2mm} $\{1,1,2\}  $ & $ \xrightarrow[\hspace{1.0cm}]{}                          $ & $ \{1,3\}   $ &:&  $ -2m(\{1,2\})             $ \\
\vspace{2mm} $\{2,2\}    $ & $ \xrightarrow[\hspace{1.0cm}]{}                          $ & $ \{2,2\}   $ &:&  $ 1                        $ \\
\vspace{2mm} $\{1,1,2\}  $ & $ \xrightarrow[\hspace{1.0cm}]{}                          $ & $ \{1,1,2\} $ &:&  $ 1                        $ \\
\vspace{2mm} $\{1,1,1,1\}  $ & $ \xrightarrow[\hspace{1.0cm}]{}                          $ & $ \{1,1,1,1\} $ &:&  $ 1                        $ \\
\end{tabular}
\label{eq:n4m}
\end{equation}
%-------------------------------------------------------------------------------------
After combining the factors from \eq{n4m} with appropriate factor from \eq{n4c}, 
as dictated by \eq{MeasFunc}, we get 
\begin{align} 
\cM^{(4)} 
&= 
	\cd(1)^4 + c(4) \Delta \cd(4)
	+ 4c(3) \cd(1) \Delta \cd(3)    [ 1- m(\{1,3\}) ]
\nonumber \\
& \qquad
	+ 3c(2)^2      \Delta \cd(2)^2  [ 1- m(\{2,2\}) ]
	+ 6c(2) \cd(1)^2 \Delta \cd(2) [ 1-  m(\{1,1,2 \}) ]
\nonumber \\
& \qquad \qquad
	- 12c(2) \cd(1)^2 \Delta \cd(2)  m(\{1,2\}) [1 -  m(\{1,3\}) ]
  \label{eq:MeasFuncn4}
\end{align}
We can write the expression for $ \cM^{(4)}$ in the form of \eq{Mnform} by substituting the expressions for $\Delta \cM^{(2)}$ and $\Delta \cM^{(3)}$
given in \eq{DeltaM2} and \eq{DeltaM3}. The result is 
\begin{align}
\cM^{(4)} = & [ \cM^{(1)} ]^4   + 6 \,  \Delta \cM^{(2)}  [ \cM^{(1)} ]^2 + 4\, \Delta\cM^{(3)} \cM^{(1)} +  \Delta \cM^{(4)} \, ,
\end{align}
where 
\begin{align}
\Delta \cM^{(4)} = &  
c(4) \Delta\cd(4) 
- 4 c(3) \cd(1) \Delta \cd(3) m\cset{1,3}
+3 c(2)^2 \Delta \cd(2)^2 \Bigl[ 1-m\cset{2,2} \Bigr]
\nn \\
\qquad \qquad &\qquad
+6 c(2)  \cd(1)^2 \Delta \cd(2) \Bigl[ 2 m\cset{1,2}m\cset{1,3}-m\cset{1,1,2} \Bigr] \,. 
\label{eq:DeltaM4}
\end{align}
Next, we give the primed version as defined under \eq{DefPrime}:
\begin{align}
\Delta \cM^{\prime(4)} &= \Delta \cM^{(4)}  - 3[ \Delta \cM^{(2)} ]^2 \\
& =   
c(4) \Delta\cd(4) 
- 4 c(3) \cd(1) \Delta \cd(3) m\cset{1,3}
- 3 c(2)^2 \Delta \cd(2)^2 m\cset{2,2}
\nn \\
\qquad \qquad &\qquad
+ 6 c(2)  \cd(1)^2 \Delta \cd(2) \Bigl[ 2 m\cset{1,2}m\cset{1,3}-m\cset{1,1,2} \Bigr] \,.
\label{eq:DeltaM4Prime}
\end{align}
Each term is associated with a map with $\ca{P}^f \to \{4\}$.

For $n = 5$, we proceed similarly.   There are seven possible partitions, each contributing to the measurement function: 
\begin{equation}
\{ \{5\}, \{1,4\}, \{2,3\}, \{1,1,3\}, \{1,2,2\},\{1,1,1,2\}, \{1,1,1,1,1\} \} \,.
\end{equation}
The factors associated with clustering are listed below.
%-------------------------------------------------------------------------------------
\begin{equation}
\begin{tabular}{ c c c  } 
   $\boldsymbol{\ca{P}^c} $   &\bf{ :} & $ \boldsymbol{N(\ca{P}^c)  \prod_{p^c \in \ca{P}^c} c(p^c) \Delta \cd (p^c) } \vspace{1mm}  $       
   \\ \vspace{-3mm} \\
\vspace{2mm} $\{5\}$       & : & $c(5)       \Delta\cd(5)               $  \\
\vspace{2mm} $\{1,4\}$     & : & $5c(4)      \cd(1)\Delta \cd(4)        $  \\
\vspace{2mm} $\{2,3\}$     & : & $10c(3)c(2) \Delta\cd(2) \Delta \cd(3) $  \\
\vspace{2mm} $\{1,1,3\}$   & : & $10c(3)     \cd(1)^2\Delta \cd(3)   $  \\
\vspace{2mm} $\{1,2,2\}$   & : & $15c(2)^2   \cd(1)\Delta \cd(2)^2   $  \\
\vspace{2mm} $\{1,1,1,2\}$ & : & $10c(2)     \cd(1)^3\Delta \cd(2)   $  \\
\vspace{2mm} $\{1,1,1,1,1\}$ & : & $ \cd(1)^5    $  \\
\end{tabular}
\end{equation}
%-------------------------------------------------------------------------------------

Next we give the factors associated with merging/clustering.  For ease of presentation, we give those maps with $\ca{P}^f =\{5\}$. 
%-------------------------------------------------------------------------------------
\begin{equation}
\begin{tabular}{c c c c c } 
 $\boldsymbol{ \phi: \ca{P}^c       }        $ & $ \xrightarrow[\hspace{1.0cm}]{}                   $ & $  \boldsymbol{\ca{P}^f     }$ &:&  $ \boldsymbol{ \ca{F} (\phi) }$ \vspace{1mm} \\  \vspace{-3mm}\\
\vspace{2mm} $\{5\}             $ & $ \xrightarrow[\hspace{1.0cm}]{}                   $ & $ \{5\}         $ &:&  $1                               $ \\
\vspace{2mm} $\{1,4\}           $ & $ \xrightarrow[\hspace{1.0cm}]{}                   $ & $ \{5\}         $ &:&  $- m(\{1,4\})                    $ \\
\vspace{2mm} $\{2,3\}           $ & $ \xrightarrow[\hspace{1.0cm}]{}                   $ & $ \{5\}         $ &:&  $- m(\{2,3\})                    $ \\
\vspace{2mm} $\{1,1,3\}         $ & $ \xrightarrow[\hspace{1.0cm}]{}                   $ & $ \{5\}         $ &:&  $- m(\{1,1,3\})                  $ \\
\vspace{2mm} $\{1,1,3\}         $ & $ \longrightarrow \quad \{1,4\} \quad    \longrightarrow             $ & $ \{5\}         $ &:&  $ 2m(\{1,3\})m(\{1,4\})           $ \\
\vspace{2mm} $\{1,2,2\}         $ & $ \xrightarrow[\hspace{1.0cm}]{}             $ & $ \{5\}         $ &:&  $-m(\{1,2,2\})                   $ \\
\vspace{2mm} $\{1,2,2\}         $ & $ \longrightarrow \quad \{2,3\} \quad    \longrightarrow             $ & $ \{5\}         $ &:&  $ 2m(\{1,2\}m(\{2,3\}))           $ \\
\vspace{2mm} $\{1,2,2\}         $ & $ \longrightarrow \quad \{1,4\} \quad    \longrightarrow           $ & $ \{5\}         $ &:&  $ m(\{2,2\}) m(\{1,4\})          $ \\
\vspace{2mm} $\{1,1,1,2\}       $ & $ \xrightarrow[\hspace{1.0cm}]{}                $ & $ \{5\}         $ &:&  $-m(\{1,1,1,2\})                 $ \\
\vspace{2mm} $\{1,1,1,2\}       $ & $ \longrightarrow \quad \{1,4\} \quad    \longrightarrow           $ & $ \{5\}         $ &:&  $ 3m(\{1,1,2\})m(\{1,4\})         $ \\
\vspace{2mm} $\{1,1,1,2\}       $ & $ \longrightarrow \hspace{2mm}  \{1,1,3\}  \hspace{2mm}  \longrightarrow           $ & $ \{5\}         $ &:&  $ 3m(\{1,2\})m(\{1,1,3\})         $ \\
\vspace{2mm} $\{1,1,1,2\}       $ & $ \to \{1,1,3\} \to \{1,4\} \to  $ & $ \{5\}         $ &:&  $-6m(\{1,2\})m(\{1,3\})m(\{1,4\}) $ \\
\vspace{2mm} $\{1,1,1,2\}       $ & $ \to \{1,1,3\} \to \{2,3\} \to  $ & $ \{5\}         $ &:&  $-3c(2)m(\{1,2\})m(\{2,3\})        $ \\
\end{tabular}
\label{eq:n5m1}
\end{equation}
%-------------------------------------------------------------------------------------
The rest of the maps are now given. 
%-------------------------------------------------------------------------------------
\begin{equation}
\begin{tabular}{c c c c c } 
 $ \boldsymbol{\phi_c: \ca{P}^c   }            $ & $ \xrightarrow[\hspace{1.0cm}]{}                   $ & $  \boldsymbol{\ca{P}^f    } $ &\bf{:}&  $\boldsymbol{\ca{F} (\phi)}$ \vspace{1mm} \\  \vspace{-3mm}\\
\vspace{2mm} $\{1,4\}           $ & $ \xrightarrow[\hspace{1.0cm}]{}                   $ & $ \{1,4\}       $ &:&  $1                               $ \\
\vspace{2mm} $\{1,1,3\}         $ & $ \xrightarrow[\hspace{1.0cm}]{}                   $ & $ \{1,4\}       $ &:&  $-2m(\{1,3\})                    $ \\
\vspace{2mm} $\{1,2,2\}         $ & $ \xrightarrow[\hspace{1.0cm}]{}                   $ & $ \{1,4\}       $ &:&  $-m(\{2,2\})                     $ \\
\vspace{2mm} $\{1,1,1,2\}       $ & $ \xrightarrow[\hspace{1.0cm}]{}                   $ & $ \{1,4\}       $ &:&  $-3m(\{1,1,2\})                  $ \\
\vspace{2mm} $\{1,1,1,2\}       $ & $\longrightarrow \quad \{1,3\} \quad    \longrightarrow              $ & $ \{1,4\}       $ &:&  $6m(\{1,2\})m(\{1,3\})           $ \\
\vspace{2mm} $\{2,3\}           $ & $ \xrightarrow[\hspace{1.0cm}]{}                   $ & $ \{2,3\}       $ &:&  $1                               $ \\
\vspace{2mm} $\{1,2,2\}         $ & $ \xrightarrow[\hspace{1.0cm}]{}                   $ & $ \{2,3\}       $ &:&  $-2m(\{1,2\})                    $ \\
\vspace{2mm} ${\bf \{1,1,1,2\}} $ & $\longrightarrow \hspace{2mm}  {\bf \{1,1,3\} } \hspace{2mm}  \longrightarrow              $ & $ {\bf \{2,3\}} $ &:&  $ \boldsymbol{3c(2)m(\{1,2\}) }          $ \\
\vspace{2mm} $\{1,1,3\}         $ & $ \xrightarrow[\hspace{1.0cm}]{}                   $ & $ \{1,1,3\}     $ &:&  $1                               $ \\
\vspace{2mm} $\{1,1,1,2\}       $ & $ \xrightarrow[\hspace{1.0cm}]{}                   $ & $ \{1,1,3\}     $ &:&  $-3m(\{1,2\})                    $ \\
\vspace{2mm} $\{1,2,2\}         $ & $ \xrightarrow[\hspace{1.0cm}]{}                   $ & $ \{1,2,2\}     $ &:&  $1                               $ \\
\vspace{2mm} $\{1,1,1,2\}       $ & $ \xrightarrow[\hspace{1.0cm}]{}                   $ & $ \{1,1,1,2\}   $ &:&  $1                               $ \\
\vspace{2mm} $\{1,1,1,1,1\}       $ & $ \xrightarrow[\hspace{1.0cm}]{}                   $ & $ \{1,1,1,1,1\}   $ &:&  $1                               $ \\
\end{tabular}
\label{eq:n5m2}
\end{equation}
%-------------------------------------------------------------------------------------
We can construct $\cM^{(5)}$ by combining these factors using \eq{MeasFunc}.  After using 
the definitions of $\Delta \cM^{(2)}$, $\Delta \cM^{(3)}$, and $\Delta \cM^{(4)}$, we get 
\begin{align}
&\cM^{(5)} =  [ \cM^{(1)} ]^5   
+ 10 \,  \Delta \cM^{(2)}  [\cM^{(1)} ]^3 
+ 10\,   \Delta\cM^{(3)} [\cM^{(1)} ]^2
+ 5\,    \Delta\cM^{(4)} \cM^{(1)}
+        \Delta \cM^{(5)} \, ,
\end{align}
where
\begin{align}
\Delta \cM^{(5)} 
 & = 
  	c(5) \Delta \cd(5)
  	- 5c(4)\cd(1) \Delta \cd(4) m(\{1,4\})  
  \nonumber \\                                              
  \qquad &		
	+10 c(3)c(2)  \Delta \cd(2) \Delta \cd(3) \Bigl[ 1 - m(\{2,3\}) \Bigr]
  \nonumber \\
  \qquad & 
  	+10c(3)  \cd(1)^2 \Delta \cd(3) 
	\Bigl[  2 m(\{1,3\})m(\{1,4\}) -  m(\{1,1,3\}) \Bigr] 
  \nonumber \\
  \qquad &   	
	  + 15c(2)^2 \cd(1) \Delta  \cd(2)^2 
	  \Bigl[ m(\{1,4\}) m(\{2,2\})  -  2m(\{1,2\}) \bigl( 1-  m(\{2,3\}) \bigr)-  m(\{1,2,2\}) \Bigr]
  \nonumber \\                                              
  \qquad &
  	+10c(2) \cd(1)^3\Delta \cd(2)                                                         
  	 \Bigl[ -  3c(2) m(\{1,2\})m(\{2,3\}) + 3 m(\{1,4\})m(\{1,1,2\})
   \nonumber \\
  \qquad & 
  	\qquad +3 m(\{1,2\}) \bigl(  -2 m(\{1,3\}) m(\{1,4\})   +m(\{1,1,3\}) \bigr)
  	- m(\{1,1,1,2\}) \Bigr]
  \nonumber \\                                              
  \qquad &	
  	 +  30c(2)^2 \cd(1)^3 \Delta \cd(2)  m(\{1,2\})   \,.
\label{eq:DeltaM5}
\end{align}
The primed version is: 
\begin{align}
\Delta \cM^{\prime(5)} &= \Delta \cM^{(5)}  - 10 \Delta \cM^{(2)} \Delta \cM^{(3)}  \nn \\
 & = 
  	c(5) \Delta \cd(5)
  	- 5c(4)\cd(1) \Delta \cd(4) m(\{1,4\})  
	 - 10 c(3)c(2)  \Delta \cd(2) \Delta \cd(3) m(\{2,3\}) 
  \nonumber \\
  \qquad & 
  	+10c(3)  \cd(1)^2 \Delta \cd(3) 
	\Bigl[  2 m(\{1,3\})m(\{1,4\}) -  m(\{1,1,3\}) \Bigr] 
  \nonumber \\
  \qquad &   	
	  + 15c(2)^2 \cd(1) \Delta  \cd(2)^2 
	  \Bigl[  m(\{1,4\}) m(\{2,2\})  +  2m(\{1,2\})  m(\{2,3\})-  m(\{1,2,2\}) \Bigr]
  \nonumber \\                                              
  \qquad &
  	+10c(2) \cd(1)^3\Delta \cd(2)                                                         
  	 \Bigl[ -  3c(2) m(\{1,2\})m(\{2,3\}) + 3 m(\{1,4\})m(\{1,1,2\})
   \nonumber \\
  \qquad & 
  	\qquad +3 m(\{1,2\}) \bigl(  -2 m(\{1,3\}) m(\{1,4\})   +m(\{1,1,3\}) \bigr)
  	- m(\{1,1,1,2\}) \Bigr]
  \nonumber \\                                              
  \qquad &	
  	 +\boldsymbol{  30c(2)^2 \cd(1)^3 \Delta \cd(2)  m(\{1,2\})  }   \,.
\label{eq:DeltaM5Prime}
\end{align}

All of the terms in $\Delta \cM^{\prime(5)}$ are associated with maps that have $\ca{P}^f = \{5\}$, 
except for the last one.
This term, whose map is shown in bold in \eq{n5m2}, is associated with a map from  $\ca{P}^c = \{1,1,1,2\}$ to $\ca{P}^f = \{2,3\}$.
This contribution is associated with two groups of gluons, one with 2 gluons and the other with 3,
rather than from clustering of all 5 gluons.  This is the first order at which clustering can occur after a merging step, 
and it is these terms which cause the difficulty.

%%%%%%%%%%%%%%%%%%%%%%%%%%%%%%%%%%%%%%%%%%%%%%%%%
\bibliography{../../jets}

\providecommand{\href}[2]{#2}\begingroup\raggedright\begin{thebibliography}{10}

\bibitem{Abdesselam:2010pt}
A.~Abdesselam, E.~Kuutmann, U.~Bitenc, G.~Brooijmans, J.~Butterworth, {\em
  et~al.}, {\it {Boosted objects: A Probe of beyond the Standard Model
  physics}},  {\em Eur.Phys.J.C} (2010)
  [\href{http://arxiv.org/abs/1012.5412}{{\tt arXiv:1012.5412}}].

\bibitem{Altheimer:2012mn}
A.~Altheimer {\em et~al.}, {\it {Jet Substructure at the Tevatron and LHC: New
  results, new tools, new benchmarks}},
  \href{http://arxiv.org/abs/1201.0008}{{\tt arXiv:1201.0008}}.

\bibitem{Catani:1991hj}
S.~Catani, Y.~L. Dokshitzer, M.~Olsson, G.~Turnock, and B.~R. Webber, {\it {New
  clustering algorithm for multi - jet cross-sections in e+ e- annihilation}},
  {\em Phys. Lett.} {\bf B269} (1991) 432--438.

\bibitem{Catani:1993hr}
S.~Catani, Y.~L. Dokshitzer, M.~H. Seymour, and B.~R. Webber, {\it
  {Longitudinally invariant K(t) clustering algorithms for hadron hadron
  collisions}},  {\em Nucl. Phys.} {\bf B406} (1993) 187--224.

\bibitem{Ellis:1993tq}
S.~D. Ellis and D.~E. Soper, {\it Successive combination jet algorithm for
  hadron collisions},  {\em Phys. Rev.} {\bf D48} (1993) 3160--3166,
  [\href{http://arxiv.org/abs/hep-ph/9305266}{{\tt hep-ph/9305266}}].

\bibitem{Dokshitzer:1997in}
Y.~L. Dokshitzer, G.~D. Leder, S.~Moretti, and B.~R. Webber, {\it Better jet
  clustering algorithms},  {\em JHEP} {\bf 08} (1997) 001,
  [\href{http://arxiv.org/abs/hep-ph/9707323}{{\tt hep-ph/9707323}}].

\bibitem{Cacciari:2008gp}
M.~Cacciari, G.~P. Salam, and G.~Soyez, {\it The anti-{$k_t$} jet clustering
  algorithm},  {\em JHEP} {\bf 04} (2008) 063,
  [\href{http://arxiv.org/abs/0802.1189}{{\tt arXiv:0802.1189}}].

\bibitem{Bartel:1986ua}
{\bf JADE} Collaboration, W.~Bartel {\em et~al.}, {\it Experimental studies on
  multijet production in {$e^+e^-$} annihilation at {PETRA} energies},  {\em Z.
  Phys.} {\bf C33} (1986) 23.

\bibitem{Brown:1990nm}
N.~Brown and W.~J. Stirling, {\it Jet cross sections at leading double
  logarithm in {$e^+e^-$} annihilation},  {\em Phys. Lett.} {\bf B252} (1990)
  657--662.

\bibitem{Bauer:2011hj}
C.~W. Bauer, N.~D. Dunn, and A.~Hornig, {\it {Subtractions for SCET Soft
  Functions}},  \href{http://arxiv.org/abs/1102.4899}{{\tt arXiv:1102.4899}}.

\bibitem{Jouttenus:2011wh}
T.~T. Jouttenus, I.~W. Stewart, F.~J. Tackmann, and W.~J. Waalewijn, {\it {The
  Soft Function for Exclusive N-Jet Production at Hadron Colliders}},  {\em
  Phys. Rev.} {\bf D83} (2011) 114030,
  [\href{http://arxiv.org/abs/1102.4344}{{\tt arXiv:1102.4344}}].

\bibitem{Ellis:2009wj}
S.~D. Ellis, A.~Hornig, C.~Lee, C.~K. Vermilion, and J.~R. Walsh, {\it
  {Consistent Factorization of Jet Observables in Exclusive Multijet
  Cross-Sections}},  {\em Phys. Lett.} {\bf B689} (2010) 82--89,
  [\href{http://arxiv.org/abs/0912.0262}{{\tt arXiv:0912.0262}}].

\bibitem{Ellis:2010rwa}
S.~D. Ellis, C.~K. Vermilion, J.~R. Walsh, A.~Hornig, and C.~Lee, {\it {Jet
  Shapes and Jet Algorithms in SCET}},  {\em JHEP} {\bf 11} (2010) 101,
  [\href{http://arxiv.org/abs/1001.0014}{{\tt arXiv:1001.0014}}].

\bibitem{Thaler:2010tr}
J.~Thaler and K.~Van~Tilburg, {\it {Identifying Boosted Objects with
  N-subjettiness}},  {\em JHEP} {\bf 03} (2011) 015,
  [\href{http://arxiv.org/abs/1011.2268}{{\tt arXiv:1011.2268}}].

\bibitem{Jankowiak:2011qa}
M.~Jankowiak and A.~J. Larkoski, {\it {Jet Substructure Without Trees}},  {\em
  JHEP} {\bf 06} (2011) 057, [\href{http://arxiv.org/abs/1104.1646}{{\tt
  arXiv:1104.1646}}].

\bibitem{Thaler:2011gf}
J.~Thaler and K.~Van~Tilburg, {\it {Maximizing Boosted Top Identification by
  Minimizing N- subjettiness}},  \href{http://arxiv.org/abs/1108.2701}{{\tt
  arXiv:1108.2701}}.

\bibitem{Walsh:2011fz}
J.~R. Walsh and S.~Zuberi, {\it {Factorization Constraints on Jet
  Substructure}},  \href{http://arxiv.org/abs/1110.5333}{{\tt
  arXiv:1110.5333}}.

\bibitem{Berger:2003iw}
C.~F. Berger, T.~Kucs, and G.~Sterman, {\it Event shape / energy flow
  correlations},  {\em Phys. Rev.} {\bf D68} (2003) 014012,
  [\href{http://arxiv.org/abs/hep-ph/0303051}{{\tt hep-ph/0303051}}].

\bibitem{Bauer:2008dt}
C.~W. Bauer, S.~Fleming, C.~Lee, and G.~Sterman, {\it Factorization of
  {$e^+e^-$} event shape distributions with hadronic final states in {Soft
  Collinear Effective Theory}},  {\em Phys. Rev.} {\bf D78} (2008) 034027,
  [\href{http://arxiv.org/abs/0801.4569}{{\tt arXiv:0801.4569}}].

\bibitem{Hornig:2009vb}
A.~Hornig, C.~Lee, and G.~Ovanesyan, {\it Effective predictions of event
  shapes: Factorized, resummed, and gapped angularity distributions},  {\em
  JHEP} {\bf 05} (2009) 122, [\href{http://arxiv.org/abs/0901.3780}{{\tt
  arXiv:0901.3780}}].

\bibitem{Stewart:2010tn}
I.~W. Stewart, F.~J. Tackmann, and W.~J. Waalewijn, {\it {N-Jettiness: An
  Inclusive Event Shape to Veto Jets}},  {\em Phys.Rev.Lett.} {\bf 105} (2010)
  092002, [\href{http://arxiv.org/abs/1004.2489}{{\tt arXiv:1004.2489}}].

\bibitem{Ellis:2007ib}
S.~Ellis, J.~Huston, K.~Hatakeyama, P.~Loch, and M.~Tonnesmann, {\it {Jets in
  hadron-hadron collisions}},  {\em Prog.Part.Nucl.Phys.} {\bf 60} (2008)
  484--551, [\href{http://arxiv.org/abs/0712.2447}{{\tt arXiv:0712.2447}}].

\bibitem{Salam:2009jx}
G.~P. Salam, {\it Towards jetography},
  \href{http://arxiv.org/abs/0906.1833}{{\tt arXiv:0906.1833}}.

\bibitem{Banfi:2005gj}
A.~Banfi and M.~Dasgupta, {\it {Problems in resumming interjet energy flows
  with $k_t$ clustering}},  {\em Phys.Lett.} {\bf B628} (2005) 49--56,
  [\href{http://arxiv.org/abs/hep-ph/0508159}{{\tt hep-ph/0508159}}].

\bibitem{Delenda:2006nf}
Y.~Delenda, R.~Appleby, M.~Dasgupta, and A.~Banfi, {\it {On QCD resummation
  with k(t) clustering}},  {\em JHEP} {\bf 0612} (2006) 044,
  [\href{http://arxiv.org/abs/hep-ph/0610242}{{\tt hep-ph/0610242}}].

\bibitem{KhelifaKerfa:2011zu}
K.~Khelifa-Kerfa, {\it {Non--global logs and clustering impact on jet mass with
  a jet veto distribution}},  \href{http://arxiv.org/abs/1111.2016}{{\tt
  arXiv:1111.2016}}.

\bibitem{Kelley:2012kj}
R.~Kelley, J.~R. Walsh, and S.~Zuberi, {\it {Abelian Non-Global Logarithms from
  Soft Gluon Clustering}},  \href{http://arxiv.org/abs/1202.2361}{{\tt
  arXiv:1202.2361}}.

\bibitem{Dasgupta:2001sh}
M.~Dasgupta and G.~P. Salam, {\it Resummation of non-global {QCD} observables},
   {\em Phys. Lett.} {\bf B512} (2001) 323--330,
  [\href{http://arxiv.org/abs/hep-ph/0104277}{{\tt hep-ph/0104277}}].

\bibitem{Dasgupta:2002bw}
M.~Dasgupta and G.~P. Salam, {\it {Accounting for coherence in interjet E(t)
  flow: A case study}},  {\em JHEP} {\bf 03} (2002) 017,
  [\href{http://arxiv.org/abs/hep-ph/0203009}{{\tt hep-ph/0203009}}].

\bibitem{Banfi:2010pa}
A.~Banfi, M.~Dasgupta, K.~Khelifa-Kerfa, and S.~Marzani, {\it {Non-global
  logarithms and jet algorithms in high-pT jet shapes}},  {\em JHEP} {\bf 08}
  (2010) 064, [\href{http://arxiv.org/abs/1004.3483}{{\tt arXiv:1004.3483}}].

\bibitem{Kelley:2011aa}
R.~Kelley, M.~D. Schwartz, R.~M. Schabinger, and H.~X. Zhu, {\it {Jet mass with
  a jet veto at two loops and the universality of non-global structure}},
  \href{http://arxiv.org/abs/1112.3343}{{\tt arXiv:1112.3343}}.

\bibitem{Kelley:2011tj}
R.~Kelley, M.~D. Schwartz, and H.~X. Zhu, {\it {Resummation of jet mass with
  and without a jet veto}},  \href{http://arxiv.org/abs/1102.0561}{{\tt
  arXiv:1102.0561}}.

\bibitem{Hornig:2011tg}
A.~Hornig, C.~Lee, J.~R. Walsh, and S.~Zuberi, {\it {Double Non-Global
  Logarithms In-N-Out of Jets}},  \href{http://arxiv.org/abs/1110.0004}{{\tt
  arXiv:1110.0004}}.

\end{thebibliography}\endgroup
%%%%%%%%%%%%%%%%%%%%%%%%%%%%%%%%%%%%%%%%%%%%%%%%%

%%%%%%%%%%%%%%%%%%%%%%%%%%%%%%%%%%%%%%%%%%%%%%%%%
\end{document}